\documentclass[aps,prd,12pt,preprint,superscriptaddress,nofootinbib,floatfix]{revtex4}
\pdfoutput=1
\usepackage{latexsym}
\usepackage{amsfonts}
\usepackage[latin1]{inputenc}
\usepackage[caption=false]{subfig}
\usepackage{graphicx}
\usepackage{mathrsfs}
\usepackage{amssymb}
\usepackage{amsmath}
\usepackage{verbatim}
\usepackage{multirow}
\usepackage{color}
\usepackage{epsfig}
\usepackage{amscd}
\usepackage{bm}
\usepackage{hyperref}

\def\D0{\slash\!\!\!\!\!\!\!\!\!\:D0}

\begin{document}

\preprint{PSI-PR-13-14}
\date{\today}

\title{Future Electron-Positron Colliders \\
and the 4-Dimensional Composite Higgs Model}

\author{D.~Barducci}\email[E-mail: ]{d.barducci@soton.ac.uk}
\affiliation{School of Physics and Astronomy, University of Southampton, Southampton SO17 1BJ, U.K.}
\author{S.~De~Curtis}\email[E-mail: ]{decurtis@fi.infn.it}
\affiliation{INFN, Sezione di Firenze, Via G. Sansone 1, 50019 Sesto Fiorentino, Italy}
\author{S.~Moretti}\email[E-mail: ]{s.moretti@soton.ac.uk}
\affiliation{School of Physics and Astronomy, University of Southampton, Southampton SO17 1BJ, U.K.}
\affiliation{Particle Physics Department, Rutherford Appleton Laboratory, Chilton, Didcot, Oxon OX11 0QX, UK}
\author{G.~M.~Pruna}\email[E-mail: ]{giovanni-marco.pruna@psi.ch}
\affiliation{Paul Scherrer Institute, CH-5232 Villigen PSI, Switzerland}

\begin{abstract}
\noindent
In this note we analyse the prospects of a future electron-positron collider in testing a particular realisation of a composite Higgs model encompassing partial compositeness, namely, the 4-Dimensional Composite Higgs Model.
We study the main Higgs production channels for three possible energy stages and different luminosity options of such a machine and confront our results to the expected experimental accuracies in the various Higgs decay channels accessible herein and, for comparison, also at the Large Hadron Collider.
\end{abstract}

\maketitle

\newpage


\section{Introduction}
\label{Sec:Intro}
\noindent

The long time effort of the physics community to understand the origin of mass has now produced an outstanding result, as the  discovery at the Large Hadron Collider (LHC) of a Higgs boson with a mass of roughly 125 GeV has finally been confirmed by both the ATLAS \cite{Aad:2012tfa} and CMS \cite{Chatrchyan:2012ufa} collaborations.
While this is an extraordinary outcome of the CERN machine operations, one of the primary questions that we need to ask now is whether this particle corresponds to the Higgs boson embedded in the Standard Model (SM) or else belongs to some
Beyond the SM (BSM) scenario.

In fact the SM suffers from a key theoretical drawback, the so-called ''hierarchy" problem, pointing out that the it could be a low energy effective theory valid only up to some cut-off energy $\Lambda$. The latter can well be at the TeV scale, hence in an energy range accessible at the LHC, so that new physics could be discovered at the CERN machine in the coming years.

For this reason many BSM scenarios, which could be revealed at the TeV scale, were proposed in the last decades, the most popular being Supersymmetry (SUSY) that solves the instability of the Higgs mass under radiative corrections by predicting new particle states in the spectrum, differing from the SM ones by half unit of spin, i.e., by postulating a symmetry between fermions and bosons \cite{Golfand:1971iw,Wess:1974tw}. The behaviour of the new SUSY particles with respect to the SM ones is such that the Higgs mass becomes stable to all orders in perturbation theory (see, e.g., \cite{Djouadi:2005gj} and references therein).

Another intriguing possibility is that the Higgs particle may be a composite state arising from some strongly interacting dynamics at a high scale instead of  being an elementary state. This will solve radically the hierarchy problem owing to compositeness form factors taming the otherwise divergent growth of the Higgs mass upon quantum effects.
Furthermore, the now measured light mass of this potentially composite Higgs state is consistent with the fact that the latter could arise as a Pseudo Nambu-Goldstone Boson (PNGB) from a particular coset of a global symmetry breaking \cite{Kaplan:1983fs,Georgi:1984af,Georgi:1984ef,Dugan:1984hq}.

Models with a composite Higgs state arising as a PNGB generally predict modifications of its couplings to both bosons and fermions of the SM \cite{Espinosa:2010vn}, hence the measurement of these quantities represents a powerful way to test the possible non-fundamental nature of the newly discovered state. Furthermore, the presence of additional particles in the spectrum of such composite Higgs models  leads to both mixing effects with the SM states as well as new Feynman diagram topologies both of which would represent a further source of deviations from the SM expectations.

The advantages of an electron-positron collider with respect to a hadron collider, e.g., the cleanliness of the environment, the precision of the measurements and the large number of Higgs bosons produced, make an analysis of the prospect of these types of machine in disentangling the fundamental from the composite nature of the Higgs boson of primary importance in view of a future decision of the physics community regarding the new generation of accelerators.

For this reason, in this note we will be analysing the generic potential of the proposed $e^+e^-$ colliders in testing a particular realisation of a composite Higgs model, the so-called 4-Dimensional Composite Higgs Model (4DCHM) of Ref.~\cite{DeCurtis:2011yx}. We will therefore borrow energy and luminosity configurations from machines prototypes such as the International Linear Collider (ILC), the Compact Linear Collider (CLIC) and the Triple Large Electron-Positron (TLEP) collider.

The plan of the paper is at follows. In the next section we describe the features of the model we have chosen. Sect.~\ref{Sec:Results} illustrates our analysis and presents our results. Sect.~\ref{Sec:Conclusions} summarises our findings.


\section{Model}
\label{Sec:Model}
\noindent

In this section we briefly introduce the salient features of the 4DCHM, that is, the specific realisation of the composite Higgs model with partial compositeness \cite{Kaplan:1991dc} that we have chosen for our analysis.

The 4DCHM is an effective low-energy Lagrangian approximation that represents an extremely deconstructed version of the Minimal Composite Higgs Model (MCHM) described in \cite{Agashe:2004rs}. It is based on the $SO(5)/SO(4)$ coset that gives four PNGBs, one of which being the physical Higgs state.
The explicit breaking of the $SO(4)$ global symmetry via linear mixing between the SM and new fields give rise to a radiative Higgs potential \emph{\`a la} Coleman-Weinberg  \cite{Coleman:1973jx} and hence to a mechanism of dynamical Electro-Weak Symmetry Breaking (EWSB).
The particular choice of the fermionic sector of the 4DCHM makes the Higgs potential Ultra-Violet (UV) finite at one loop and, remarkably, for a natural choice of the model parameters, it has the correct form to provide EWSB \cite{DeCurtis:2011yx} and a Higgs boson mass compatible with the experimental value of the new state discovered at the LHC.

The extra $SO(5)\otimes U(1)_X$ gauge group present in the model implies 11 new vectorial degrees of freedom among which 5 are neutral and $3+3$ are charged gauge bosons, collectively called $Z^\prime$ and $W^\prime$, respectively. The choice of the four fundamental representations of $SO(5)\otimes~U(1)_X$ (called $\Psi_{T,B}$ and $\tilde \Psi_{T,B}$) that build up the fermionic sector gives 20 new fermions, 16 with standard electric charge, $+2/3$ and $-1/3$, denoted collectively as $t^\prime$ and $b^\prime$, respectively, and 4 with exotic charges, $+5/3$ and $-4/3$, denoted collectively as $\tilde t^\prime$ and $\tilde b^\prime$, respectively. In our notation, the new particles present in the 4DCHM in addition to the SM ones are\footnote{An increasing number labelling the 4DCHM bosonic and fermionic states corresponds to their increasing masses.} then the following:
\begin{itemize}
\item Neutral gauge bosons: $Z_i$ (with $i=1, ... 5$)
\item Charged gauge bosons: $W^\pm_i$ (with $i=1, ... 3$)
\item Charged $+2/3$ fermions: $t_i$  (with $i=1, ...8$)
\item Charged $-1/3$ fermions: $b_i$ (with $i=1,...8$)
\item Charged $+5/3$ fermions: $\tilde{t}^\prime_{i}$ (with $i=1,2$)
\item Charged $-4/3$ fermions: $\tilde{b}^\prime_{i}$ (with $i=1,2$)
\end{itemize}

Together with the SM matter and gauge fields and neglecting the $SU(3)_C$ part that is left untouched with respect to the SM (besides the fact that also the new spin $1/2$ resonances carries colour charge and so the corresponding QCD terms ought to be added) the 4DCHM is described by the following Lagrangian
\begin{equation}
 \begin{split}
 &\mathcal{L}=\mathcal{L}_{gauge}+\mathcal{L}_{ferm},\\
 &\mathcal{L}_{gauge}=\frac{f_1^2}{4}Tr|D_{\mu}\Omega_1|^2+\frac{f_2^2}{2}(D_{\mu}\Phi)(D_{\mu}\Phi)^T-
  \frac{1}{4}\rho_{\mu\nu}^{\tilde A}\rho^{{\tilde A}\mu\nu}-\frac{1}{4}F_{\mu\nu}^{\tilde W} F^{{\tilde W}\mu\nu}, \\ 
 &\mathcal{L}_{ferm}=\mathcal{L}_{ferm}^{el}+ (\Delta_{t_L}\bar{q}^{el}_L\Omega_1\Psi_T+\Delta_{t_R}\bar{t}^{el}_R\Omega_1\Psi_{\tilde{T}}+h.c.)\\
  &+\bar{\Psi}_T(i\hat{D}^{\tilde{A}}-m_*)\Psi_T+\bar{\Psi}_{\tilde{T}}(i\hat{D}^{\tilde{A}}-m_*)\Psi_{\tilde{T}}\\
  &-(Y_T\bar{\Psi}_{T,L}\Phi^T\Phi\Psi_{\tilde{T},R}+M_{Y_T}\bar{\Psi}_{T,L}\Psi_{\tilde{T},R}+h.c.)\\
  &+(T\rightarrow B).
 \end{split}
 \label{L4DCHM}
\end{equation}
In  $\mathcal{L}_{gauge}$ the covariant derivatives with respect to the SM $\tilde W$ fields and the extra $\tilde A$ fields, with coupling $g_0$ and $g_{\rho}$, are given by
 \begin{equation}
 \begin{split}
  &D^{\mu}\Omega_1=\partial^{\mu}\Omega_1-i g_{0}\tilde{W}^\mu\Omega_1+i g_{\rho}\Omega_1\tilde{A}^\mu,\\
  &D_{\mu}\Phi=\partial_{\mu}\Phi-i g_{\rho}\tilde{A}_\mu\Phi.
  \end{split}
 \end{equation}
The link fields $\Omega_n$ in the unitary gauge are given by
\begin{equation}
 \Omega_n={\bf{1}}+i\frac{s_n}{h}\Pi+\frac{c_n-1}{h^2}\Pi^2,~~~s_n=\sin(f h/f_n^2),~~~c_n=\cos(f h/f_n^2),  ~~~h=\sqrt{h^{\hat{a}}h^{\hat{a}}},
\end{equation}
where $\Pi=\sqrt{2} h^{\hat{a}}T^{\hat{a}}$ is the PNGB matrix and the $T^{\hat{a}}$'s are the $SO(5)/SO(4)$ broken generators, with $\hat{a}=1,2,3,4$.

The field $\Phi$ is a vector of $SO(5)$  that describes the spontaneous symmetry breaking of $SO(5)\otimes U(1)_X \rightarrow SO(4)\otimes U(1)_X$ and is defined as
\begin{equation}
\Phi=\phi_0\Omega_2^T \quad \text{where} \quad \phi_0^i=\delta^{i5}.
\end{equation}

The $f_{i}$'s are the link coupling constants and $f$ the strong sector scale (or `compositeness' scale), which  are  related through
\begin{equation}
\sum_{n=1}^2\frac{1}{f_n^2}=\frac{1}{f^2}
\end{equation}
and the choice $f_1=f_2=\sqrt{2}~f$ has then been made.

The fermionic Lagrangian $\mathcal{L}_{ferm}$ contains kinetic terms, mass ($m_*$) and interaction ($Y_{T,B}, m_{Y_{T,B}}$) parameters between the extra fermions and also mixing parameters ($\Delta_{t,b/L,R}$) between the extra matter and the SM content. $\mathcal{L}^{el}_{ferm}$ describes the kinetic terms of the SM fermions.


\section{Results}
\label{Sec:Results}
\noindent

Being our goal the study of this particular composite Higgs framework in its completeness and its detailed analysis with its full particle spectrum included, we have chosen to implement the 4DCHM into numerical tools that allow one to perform dedicated analyses up to event generation.
Our simulations have been mainly performed with the help of the CalcHEP package \cite{Belyaev:2012qa} in which the model under discussion had been previously implemented via the LanHEP tool~\cite{Semenov:2010qt}, as already illustrated in some detail in Refs.~\cite{Barducci:2012kk,Barducci:2013wjc}.
Moreover, since CalcHEP only  allows by default the analysis of only tree-level processes, we have also added by hand the one-loop $Hgg$, $H\gamma\gamma$
and $H\gamma Z$ vertices (again computed with the full model particle spectrum inside them).

A feature specific to future $e^+e^-$ colliders is the presence of Initial State Radiation (ISR) and Beamstrahlung. For the former, CalcHEP implements the Jadach, Skrzypek and Ward expressions of Refs.~\cite{Jadach:1988gb,Skrzypek:1990qs}. Regarding the latter, we adopted the parametrisation specified for the ILC project in~\cite{Behnke:2013xla}, that is\footnote{While fine details of the emerging electron and positron spectra may be different in other parametrisations (see \cite{Aicheler:2012bya} for CLIC and \cite{Gomez-Ceballos:2013zzn} for TLEP),
we can confirm that the gross features of the ensuing 4DCHM phenomenology are generically captured
by the present one.},
\begin{itemize}
\item beam size $(x+y)$: $645.7$ nm,
\item bunch length: $300$ $\mu$m,
\item bunch population: $2\cdot10^{10}$.
\end{itemize}

\begin{figure}[htb]
\centering
\includegraphics[width=0.5\linewidth]{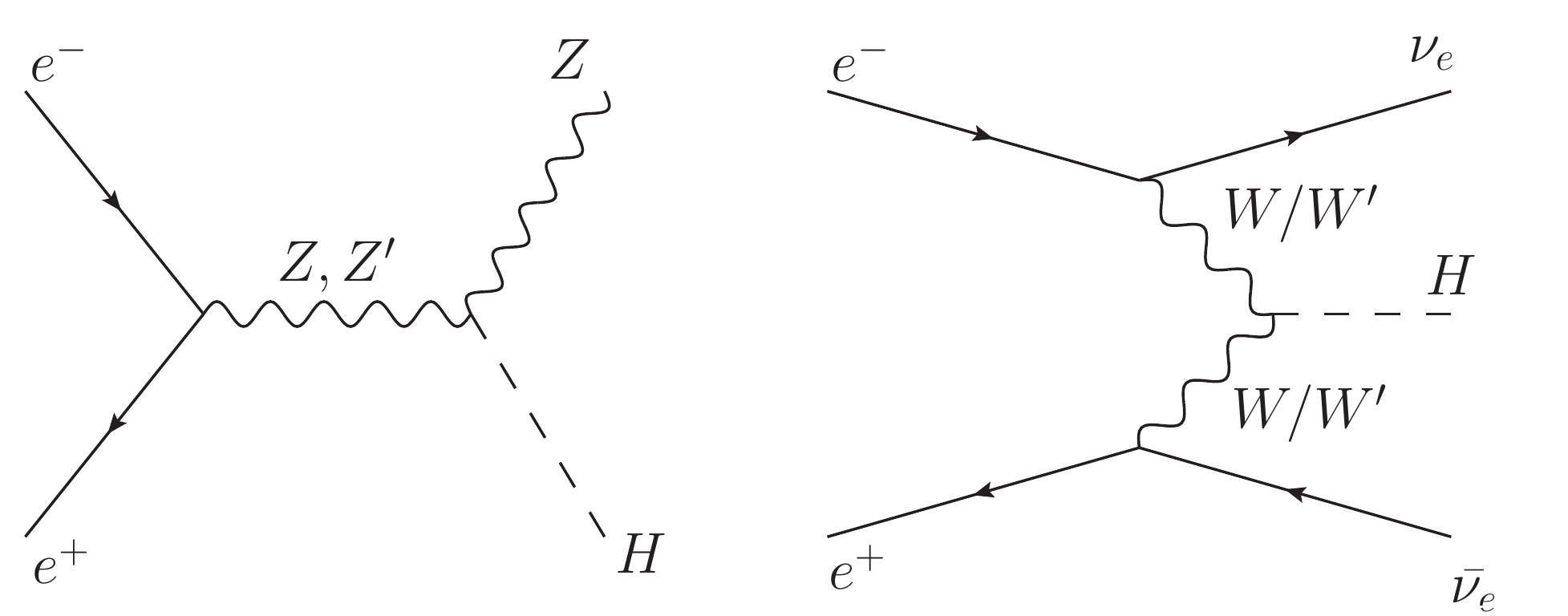}{}
\caption{Representative Feynman diagrams for the HS (left) and VBF (right) single Higgs production processes at an electron-positron collider. Extra gauge bosons, $Z^\prime$ and $W^\prime$, can also be exchanged.}
\label{fig:feynZHWW}
\end{figure}

\begin{figure}[h!]
\centering
\includegraphics[width=0.25\linewidth]{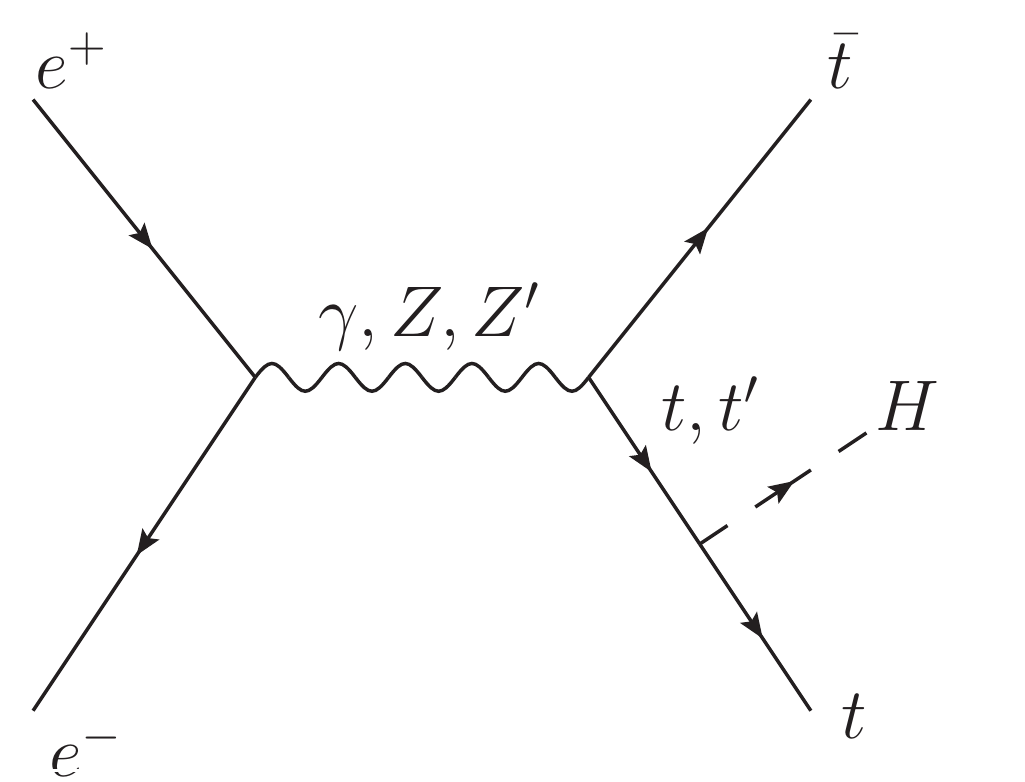}{}
\caption{Representative Feynman diagrams for single Higgs production in association with $t\bar t$ pairs  at an electron-positron collider. Extra gauge bosons, $Z^\prime$, and/or top partners, $t^\prime$, can also be exchanged.}
\label{fig:feynHtt}
\end{figure}

\begin{figure}[htb]
\centering
\includegraphics[width=0.8\linewidth]{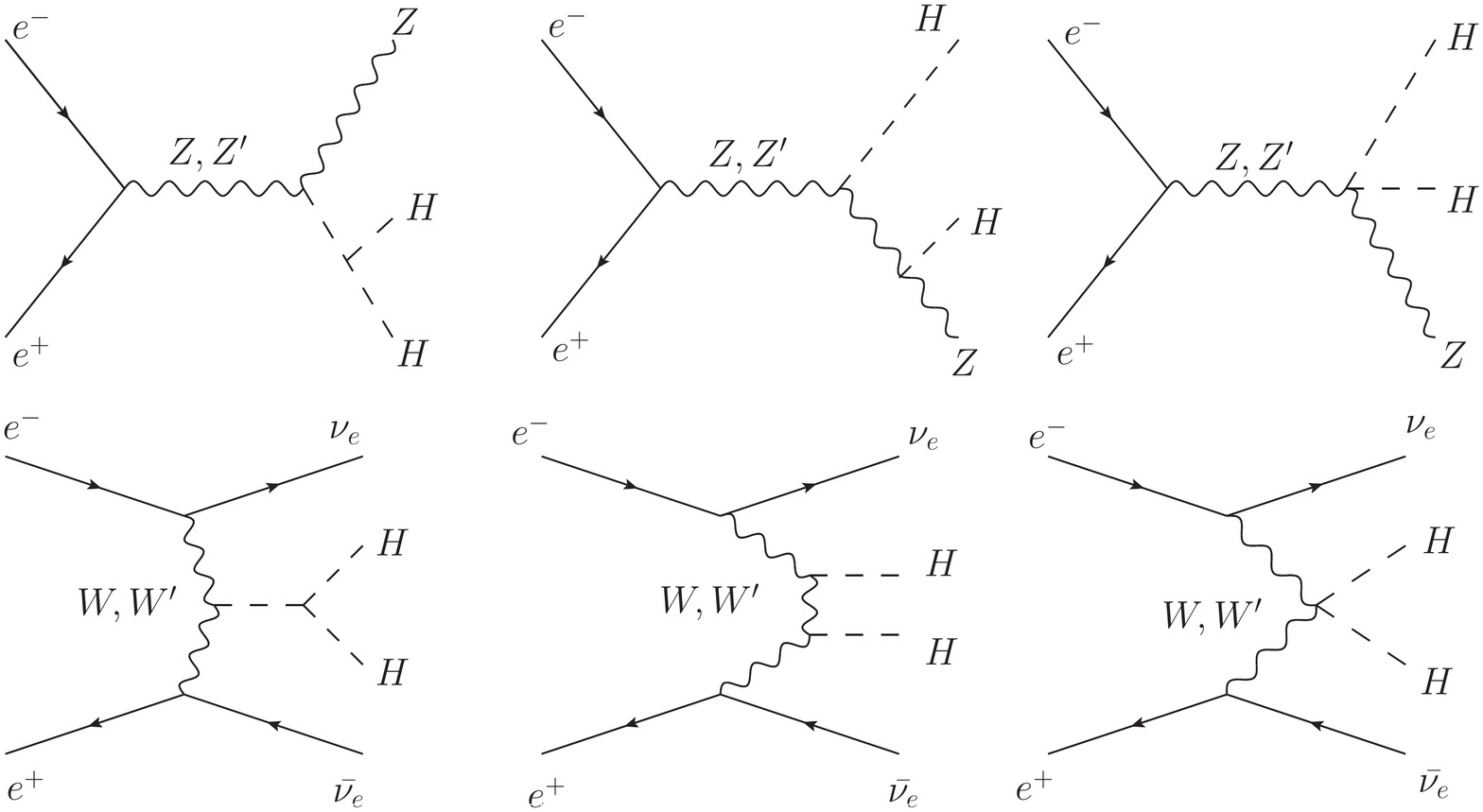}{}
\caption{Representative Feynman diagrams for double Higgs production via HS (top) and via VBF (bottom) at an electron-positron collider. Extra gauge bosons,
$W'$ and $Z'$, can also be exchanged.}
\label{fig:feynZHH}
\end{figure}

We will be considering throughout three values for the Centre-of-Mass (CM) energy, which are standard benchmark energies for a future $e^+e^-$ collider: $250$ GeV, $500$ GeV and $1$ TeV. 
Within the 4DCHM, we have studied the phenomenology of a Higgs boson obtained via the standard production mechanisms of lepton colliders: i.e., Higgs-Strahlung (HS) from neutral vector bosons and Vector Boson Fusion (VBF) (the latter primarily onset by an off-shell charged vector boson pairs),  as per Feynman diagram shown in Fig.~\ref{fig:feynZHWW}. 
Furthermore, we have focused the analysis on two other Higgs processes, which are the specialities of future electron-positron machines: i.e., Higgs production in association with top-antitop quark pairs (see Fig.~\ref{fig:feynHtt}) and double Higgs production, considering both the case of a second Higgs boson emitted in the HS channel, $e^+ e^-\to Z H H$, and the one via VBF, $e^+ e^-\to \nu_e\bar\nu_e H H$ (see Fig.~\ref{fig:feynZHH}).  
When combining production cross sections and decay Branching Ratios (BRs), our simulated data have been always related to the experimental accuracies presented in Refs.~\cite{Peskin:2012we,Asner:2013psa,Baer:2013cma}. 
There and here, we will indicate the production cross section with $\sigma(ZH)$ for HS, $\sigma(WW)$ for VBF, $\sigma(ttH)$ for Higgs production in association with $t\bar t$ pairs,  $\sigma({ZHH})$ for double HS and $\sigma(\nu \bar\nu H H)$ for double Higgs production via VBF. In keeping with the aforementioned references, we have always assumed a luminosity of $250$/$500$/$1000$ fb$^{-1}$ in correspondence to an energy of $250$/$500$/$1000$ GeV.

In the following (sub)sections we will present several results concerning the studies of the aforementioned Higgs production processes organised as follows. Initially we will confirm the results that in the so-called ``decoupling limit'' (whereby, essentially, the additional  gauge bosons and fermions present in the 4DCHM particle spectrum are made sufficiently  heavy)
 sizable deviations from the SM couplings of the discovered Higgs object are possible and may indicate indirect evidence of a composite Higgs sector.
However, we will eventually also argue that such decoupling does not really onset yet for values of the 
new particle masses pertaining to the 4DCHM which are still compatible with current experimental data: i.e., following the LHC measurement of the light Higgs boson mass and the constraints from the Electro-Weak Precision Tests (EWPTs) from LEP, SLC and Tevatron\footnote{In particular,
 the naive tree-level contribution of the new spin-1 particles to the $S$ parameter leads to a lower bound on their mass around 3 TeV. However, new positive contributions to the $T$ parameter can relax this constraint down to 2 TeV \cite{Grojean:2013qca,Contino:2013gna}.}, primarily impinging on the fermionic and gauge sector of the 4DCHM respectively, as well as the negative results from Tevatron and LHC direct searches for new gauge bosons and fermions. 
That is, we will show that genuine 4DCHM effects cannot be relegated to a simple rescaling of the relevant Higgs couplings, as the presence of $W', Z'$ and/or $t'$ propagator effects in all the production processes analysed cannot generally be neglected.
This will in fact be shown to be a relevant effect for HS Higgs production and production in association with $t\bar t$ 
pairs, albeit less so for single Higgs production via VBF and both double Higgs production channels, via HS and via VBF, the latter two being also limited in a phenomenological analysis by the poor experimental accuracies which are expected.

In essence, to generalise our findings, quantitative studies of Higgs boson 
phenomenology in composite Higgs models at future electron-positron colliders cannot ignore effects from realistic mass spectra. Conversely, studies in the decoupling scenario remain valid so long that such mass spectra are made significantly heavy. 
  
\subsection{Decoupling limit in the 4DCHM}
\label{Sec:decoup}

In recent years, a popular approach introduced by \cite{Buchmuller:1985jz} (for a more recent review, see also \cite{AguilarSaavedra:2010zi,Grzadkowski:2010es}) in order to parametrise effects of generic New Physics (NP), has been adapted to the study of Higgs couplings (see, e.g., \cite{Giudice:2007fh,Contino:2013kra,Brivio:2013pma})  in generic decoupling 
scenarios. The rationale is as follows. It is commonly accepted that any effects arising from NP (at some scale $\Lambda_{\rm NP}\gg \Lambda_{\rm EW}$) can be embedded in 
the SM via 6-dimensional effective operators (thereby assuming that the new physical fields are integrated away), realising the aforementioned decoupling limit, so long that the probing energy is around the EW scale. Then, after a specific theory is chosen, the coefficients of such effective operators can be expressed in terms of coupling modifications of the SM interactions.

Now, focusing onto the 4DCHM as an illustrative example of composite Higgs scenarios with additionally heavy matter and gauge fields, the modifications of the Higgs couplings arise via three effects:
\begin{enumerate}
\item the non-linear realisation of the Goldstone symmetry;
\item the mixing between SM and extra particles due to the partial compositeness mechanism;
\item the possibility that, in both tree-level and loop-induced processes, exchange of new particles can occur.
\end{enumerate}
In the aforementioned effective approach, the first two effects are normally  captured  while the latter is not whenever the masses of the additional states are close to $\Lambda_{\rm NP}$, as it is naturally the case. In fact,
sometimes, in current literature, even the second effect is not accounted for, so that one achieves a true decoupling of all NP, i.e., all the non-SM particle effects in the model spectrum are neglected, and the analysis consists in the trivial rescaling of the Higgs couplings due to its Goldstone nature. 
In particular, defining  $\xi=v^2/f^2$ and assuming the case of the 4DCHM (i.e., a $SO(5)/SO(4)$ coset with extra fermions in  fundamental representations of $SO(5)$), this rescaling values, for vector bosons $V=W/Z$ and SM fermions $f$, are \cite{Espinosa:2010vn}
\begin{equation}
\frac{g_{HVV}^{\rm SM}}{g_{HVV}^{\rm 4DCHM}} = \sqrt{1-\xi}, \quad \quad \quad \frac{g_{Hff}^{\rm SM}}{g_{Hff}^{\rm 4DCHM}}=\frac{1-2 \xi}{\sqrt{1-\xi}}.
\end{equation}
In such a regime of decoupling we can also compute the values of the so-called Higgs signal strengths,
defined as 
\begin{equation}\label{mu}
\mu_{i}=\frac{\sigma(e^+e^-\rightarrow H X)_{\rm 4DCHM}{\rm BR}(H\rightarrow i )_{\rm 4DCHM}}
                                          {\sigma(e^+e^-\rightarrow H X)_{\rm SM       }{\rm BR}(H\rightarrow i)_{\rm SM}        },
\end{equation}
where $X$ represents anything produced in association with the Higgs boson (e.g., $Z$ in the case of HS and $\nu_e\bar\nu_e$ or $e^+e^-$ in the case of VBF) and $i$ simply labels a possible final state of the Higgs boson decay (for example $b\bar b$, $\gamma\gamma$, $WW$, $ZZ$), and relate a measurement  of the various $\mu_i$'s to the model parameter $f$.

\begin{figure}[t!]
\centering
\includegraphics[width=1.0\linewidth]{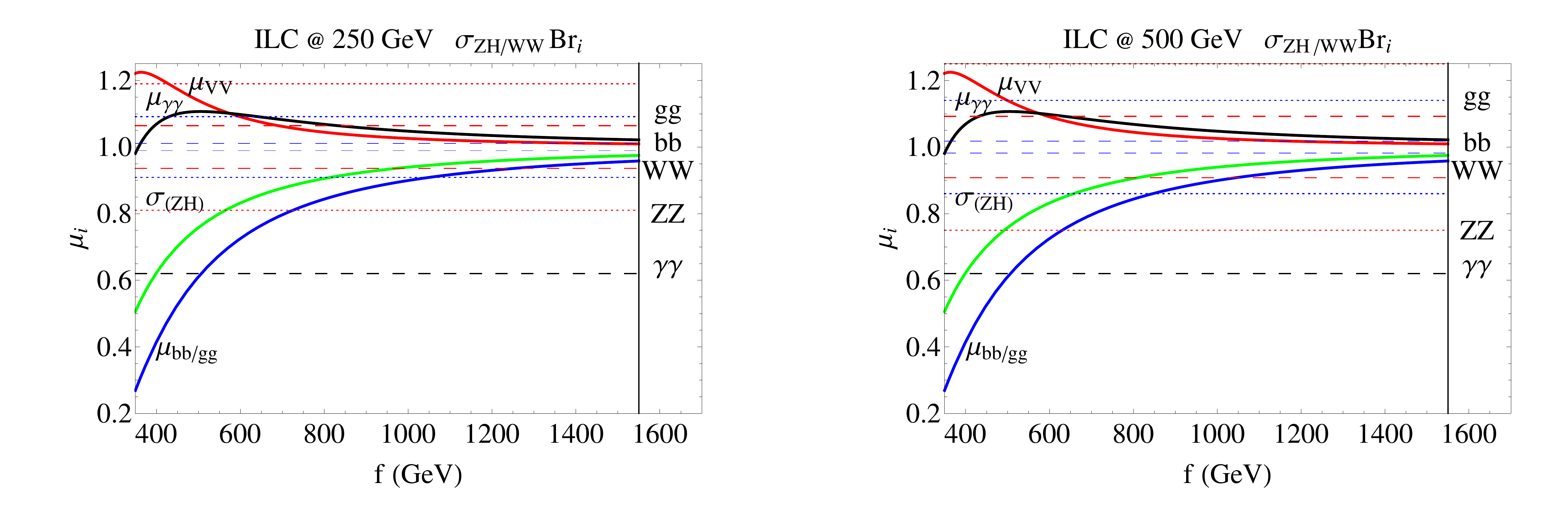}{}
\caption{Higgs signal strengths $\mu_i$ for $i=VV(=WW/ZZ)$ (red), $\gamma\gamma$ (black) and $b \bar b/gg$ (blue) as a function of the model scale $f$. In green is given the ratio of the inclusive $\sigma({ZH})$ cross section with respect to the SM value. Horizontal lines represent the expected experimental accuracies for cross section times BR measurements at the 250 
GeV (left) and 500 GeV (right) stage of a future $e^+e^-$ collider, according to Refs.~\cite{Peskin:2012we,Baer:2013cma}: $WW$ (red dashed), $ZZ$ (red dotted), $\gamma\gamma$ (black dashed), 
 $b \bar b$ (blue dashed) and $gg$ (blue dotted).}
\label{fig:XiAnLC}
\end{figure}

An example of this is found in Fig.~\ref{fig:XiAnLC}, limited for illustration purposes to the case of the HS process, which indeed shows that there exists sensitivity at future $e^+e^-$ machines to the compositeness scale $f$. 
Further exercises can be performed along similar lines and various such statements can be finally made for a variety of production modes and decay channels as recently done, e.g., in Ref.~\cite{Contino:2013gna}.

However, in the light of our previous discussion, a legitimate question arises: does the decoupling limit represent a good approximation for interpreting the phenomenological implication of the 4DCHM? The answer is ``generally not''.
In the following, we will set out to prove this by studying the impact of the 4DCHM particle spectrum onto the main Higgs production and decay channels for customary values of the NP scale.

We have started by performing a parameter scan\footnote{For the details of the latter, including the ranges adopted in modelling the parameter space as well as the experimental constraints implemented, see Refs.~\cite{Barducci:2012kk,Barducci:2013wjc}.
However, as an update on those scans, we have here excluded from our analysis points in which the mass of the SM quark partners $t',b',\tilde t',\tilde b'$ is lighter than $600~$GeV, this to take into account limits on the direct search of heavy quarks coming from the LHC experiments, see for example \cite{DeSimone:2012fs}.} for various benchmark points of our model. We have found that the largest deviations from the SM expectations arise, naturally, for the lowest possible values of $M_{Z'},M_{W'}\approx f g_\rho$ and,
 correspondingly, for the lowest choices of the  compositeness scale $f$. We have considered $f$ in the TeV range and fixed the $g_\rho$ parameter in order to have $M_{Z'},M_{W'}\approx 2$ TeV, which indeed represents an allowed configuration from both EWPTs (see, e.g., Fig.~1  of \cite{Contino:2013gna}) and direct $Z',W'$ searches (see Ref.~\cite{Barducci:2012as}). 
Furthermore, notice that keeping the mass of the extra neutral vector resonance fixed renders  clearer the dependence of the observable onto the compositeness scale $f$.

In order to separate (factorisable) rescaling effects, due to both the non-linear  realisation of the Goldstone symmetry and  the mixing between SM and extra particles, from the ones due
to the additional propagators, we have introduced the $R$ and $\Delta$ parameters for the inclusive HS production cross section as follows:
\begin{eqnarray}\label{mu_definition}
R=\frac{\sigma(ZH)_{\rm 4DCHM}}{\sigma(ZH)_{\rm SM}}, \qquad \Delta=R-\kappa^2_{HZZ}, \qquad \kappa_{HZZ}=\frac{g_{HZZ}^{\rm 4DCHM}}{g^{\rm SM}_{HZZ}}.
\end{eqnarray}
Then, by numerical computation, we have proven that, if the new class of neutral gauge bosons are completely stripped off the calculations, $R$ tends to $
\kappa^2_{HZZ}$ (equivalently, $\Delta$ tends to $0$) with a negligible deviation $\sim 0.01\%$ due to a slight shift in the $C_V$ and $C_A$ couplings of the SM-like $Z$ to the fermions (herein, $e^+e^-$),
due to the aforementioned mixing.
Since HS is generically the
most useful process from which deviations of the Higgs couplings from the SM values can be extracted, we are essentially making the generic statement that, even when the CM energy of the 
collider is below the scale of NP, $f$ in this case, where the additional boson and fermion masses of the 4DCHM naturally tend to cluster, the HS cross section is always affected by propagator effects (see also \cite{Conley:2005et} and \cite{Hartling:2012ss} for studies in the context of the Littlest Higgs and Minimal Composite Higgs models, respectively).
\begin{figure}[htb]
\centering
\includegraphics[width=0.32\linewidth]{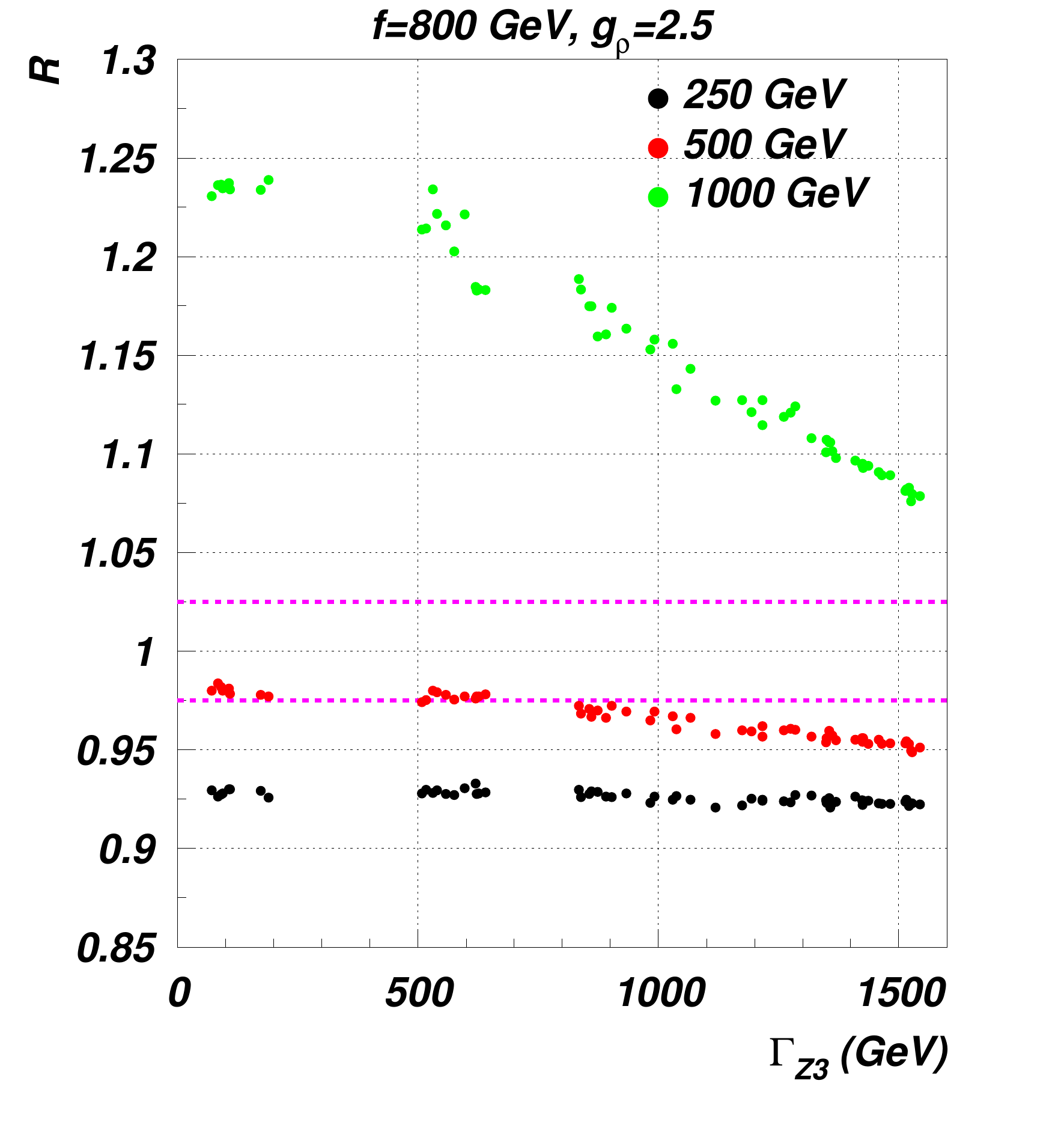}{}
\includegraphics[width=0.32\linewidth]{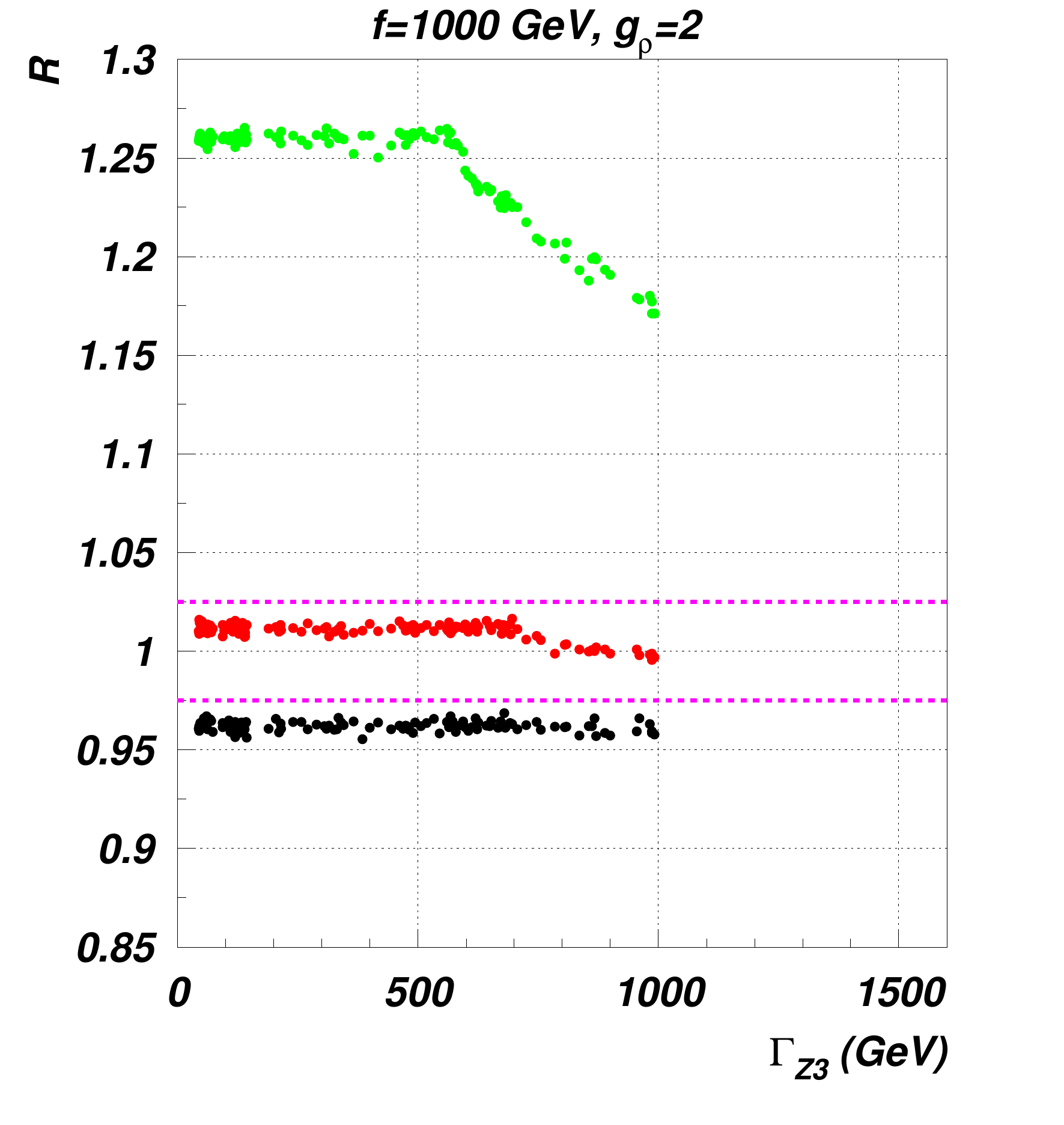}{}
\includegraphics[width=0.32\linewidth]{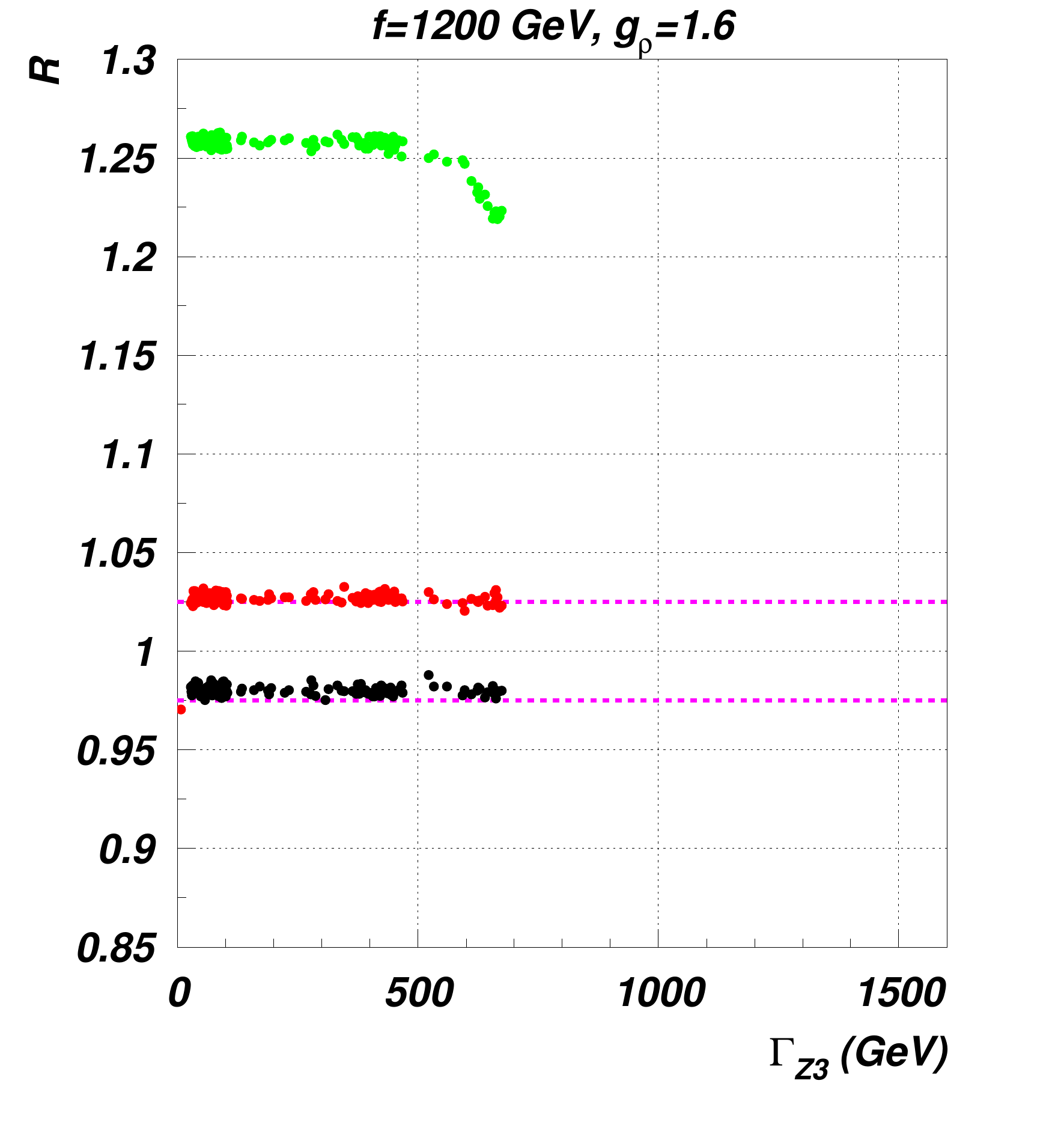}{} \\
\includegraphics[width=0.32\linewidth]{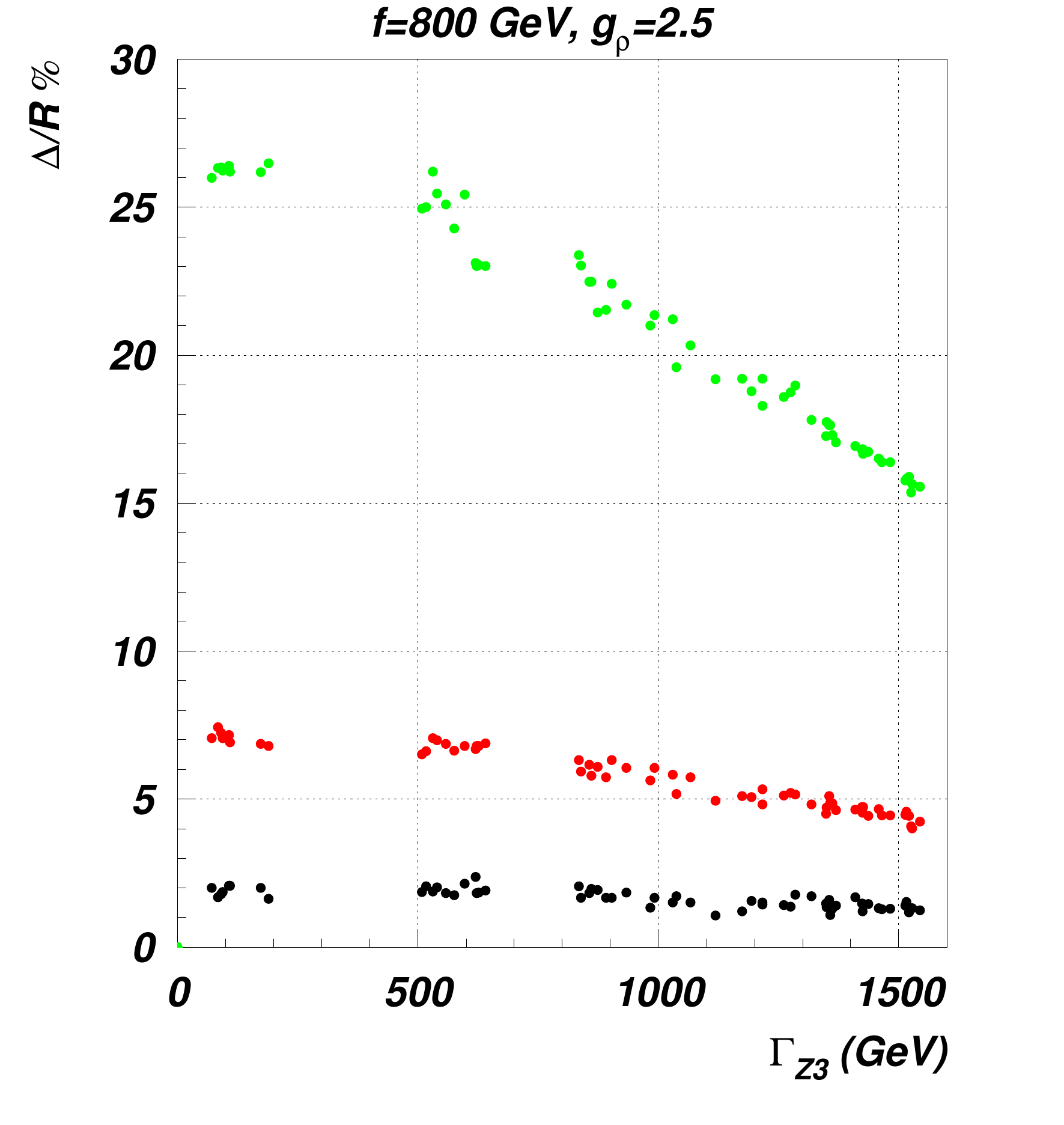}{}
\includegraphics[width=0.32\linewidth]{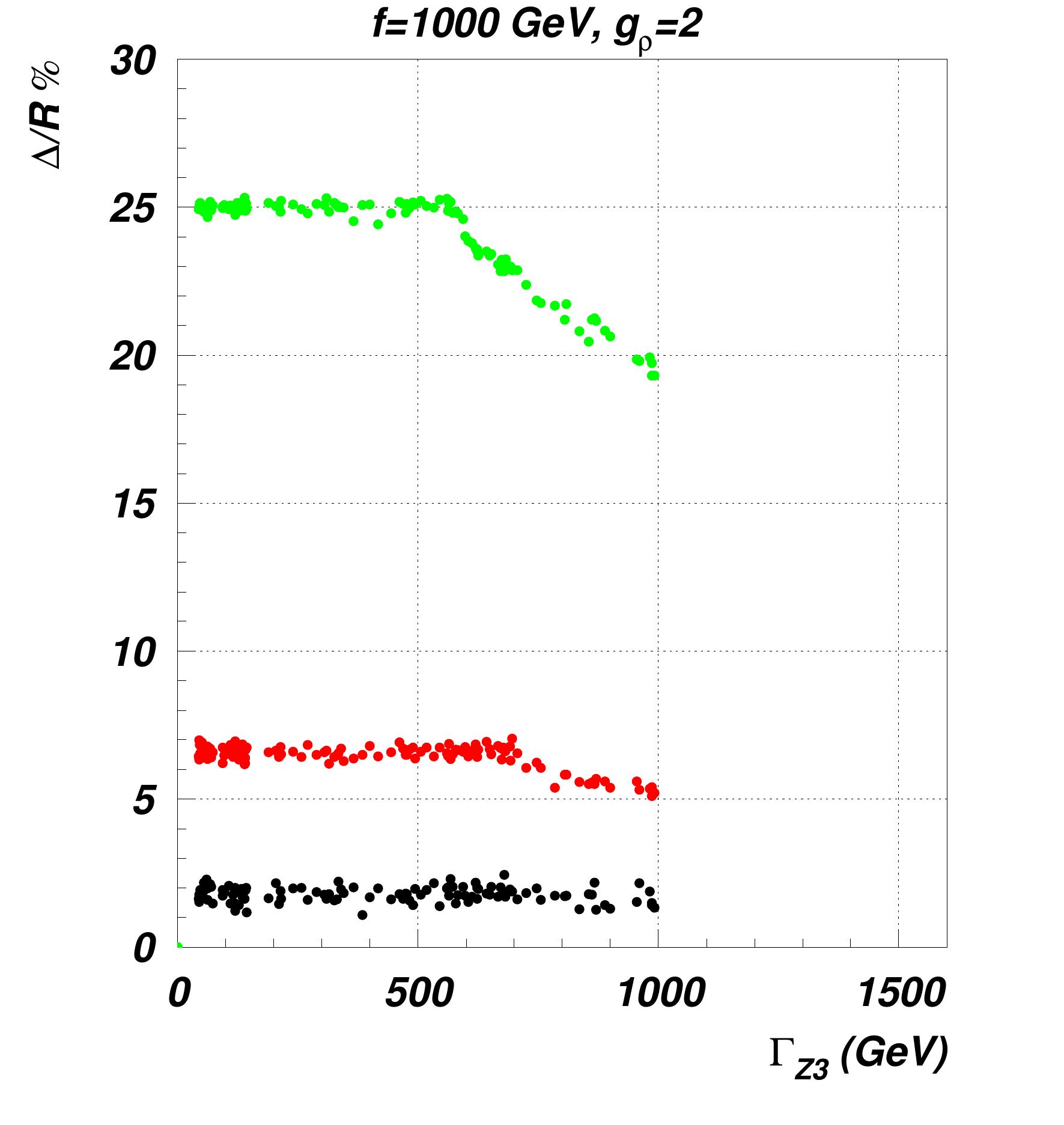}{}
\includegraphics[width=0.32\linewidth]{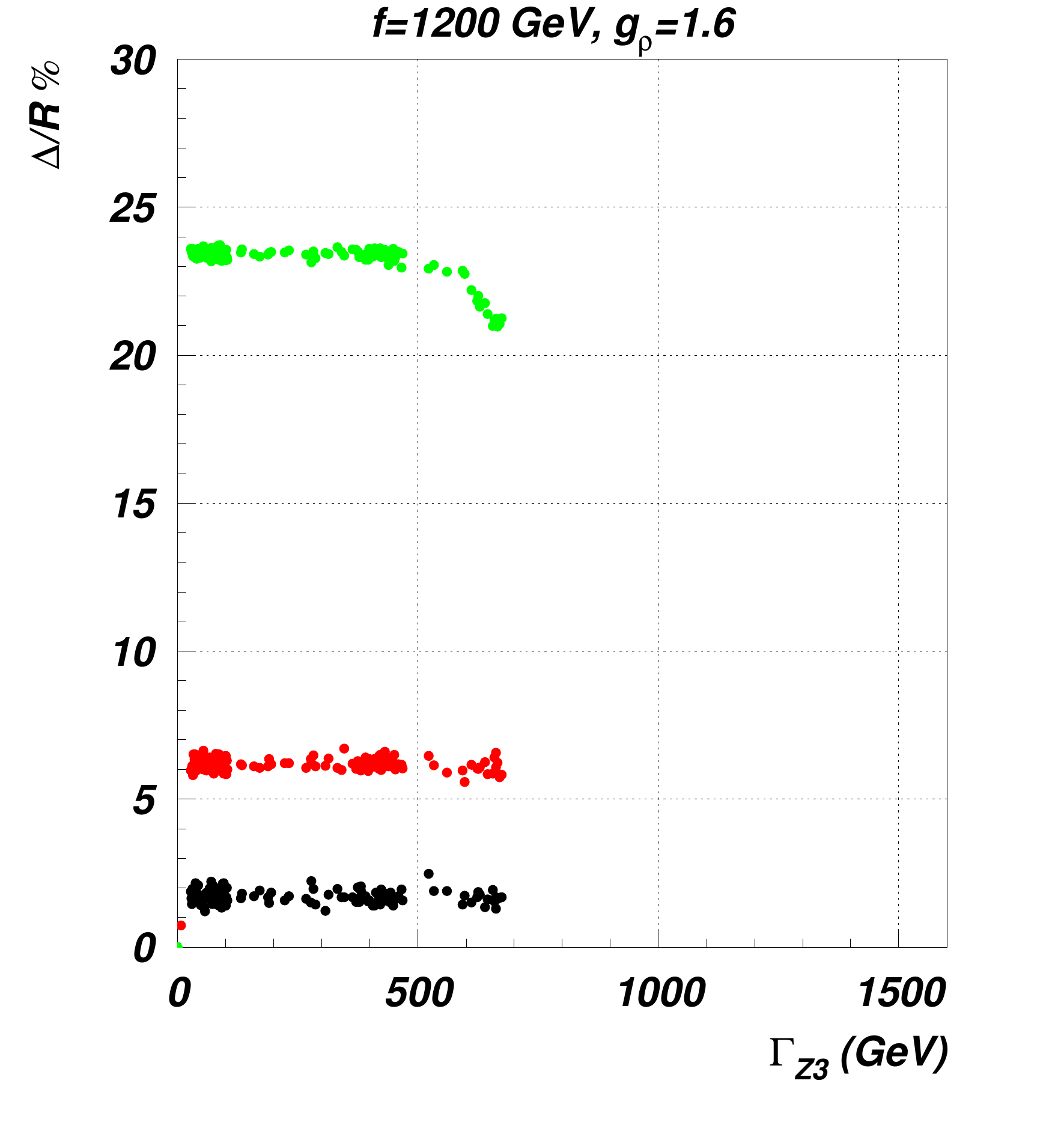}{}
\caption{The $R$ and $\Delta$ quantities defined in eq.~(\ref{mu_definition}) plotted against the width of the $Z_3$ resonance for the benchmark points with $f=800$ GeV and $g_\rho=2.5$ (left), $f=1000$ GeV and $g_\rho=2$ (centre), $f=1200$ GeV and $g_\rho=1.6$ (right). The dotted purple line represents the experimental precision in determining $R$, according to Refs.~\cite{Peskin:2012we,Baer:2013cma}.}
\label{fig:HS}
\end{figure}
This is well illustrated by Fig.~\ref{fig:HS}, where we quantify the $R$ and $\Delta$ parameters for three of our benchmarks choices ($f=800~$GeV, $f=1000~$GeV and $f=1200~$GeV, corresponding to a value of the $\xi$ parameter of 0.1, 0.06 and 0.04, respectively) as a function of the total width of the dominant\footnote{We explicitly verified that all the features of 
the $Z_2$ and $Z_3$ contributions are the same, with the exception of the couplings involved: the $Z_3$ one to leptons is about twice the $Z_2$ one to the same particles. $Z_1$ and $Z_4$ do not couple to leptons and are thus irrelevant in this process. $Z_5$ is always much heavier.} extra-vectorial contribution, i.e., $\Gamma_{Z_3}$, for the
 three customary values of CM energy.
 The rescaling factors are $\kappa^2_{HZZ}\approx0.91$ for 
$f=800~$GeV, $\kappa^2_{HZZ}\approx0.94$ for $f=1000~$GeV and $\kappa^2_{HZZ}\approx0.96$ for $f=1200~$GeV.
The slopes present in the plots, more noticeable the larger the CM energy, make it clear 
that one is in presence of width dependent propagator effects. In fact, the trend of $R$ (or equivalently $\Delta$) is almost constant but, from some threshold on, it decreases with $\Gamma_{Z_3}$ (somewhat linearly) reflecting the nature of the interference contribution that is proportional to $1/\Gamma_{Z_3}$ when the CM energy is 
smaller than the $Z_3$ masses involved (modulus some dilution induced by the $Z_2$ state, {which is however responsible for the interference in the region where $\Gamma_{Z_3}$ is small).}

The deviations from the SM limit span from $\sim 2\%$ when $\sqrt{s}=250$ GeV up to $\sim 25\%$ when $\sqrt{s}=1$ TeV. We have also verified that the effect is completely due to 
the constructive interference term arising from the SM-like $Z$ resonance and the $Z_2+Z_3$ contributions, with $Z_3$ being dominant among the two extra vectors, as already mentioned. A feature of such interference effect is that the $R$ values are always above the expected ``reduction''  onsetting in the decoupling limit,  to the point that, at $\sqrt{s}=1$ TeV 
for the three benchmarks and even at $\sqrt{s}=500~$GeV for $f=1000~$GeV and $f=1200~$GeV, the $R$ value is above $1$, which is never expected if the new resonances are totally decoupled.

We have then verified that choosing a bigger $M_{Z'},M_{W'}$  will render the effects of the extra gauge bosons smaller
until one reaches the limit in which they are completely decoupled from the theory, thereby recovering the results of \cite{Contino:2013gna} although, for example, even for $M_{Z'}=3$~TeV, the effects due to the aforementioned interference are still present and lead to deviations up to $\sim 10\%$ with respect to the scaling of the couplings, when $\sqrt{s}=1$ TeV.

As a conclusion then, we believe that a complete study of composite Higgs models via the HS process should also take into account the possibility of non-decoupled spectra for the extra resonances (in the
context of composite Higgs scenarios the effects of resonances, although in different processes, has
already been analysed in \cite{Contino:2011np}). 

For completeness, we have also analysed the case of  VBF,  where we have established that such interference effect is taking place but with inverted sign and with a contribution that is one order of magnitude lower than in the HS case. The largest effect  is again seen at $\sqrt{s}=1$ TeV, yielding a destructive interference of $\sim 2-3\%$.

\subsection{Away from the decoupling limit in the 4DCHM}

\subsubsection{Higgs couplings analysis at the LHC}
\label{Sec:LHC}

\begin{figure}[h!]
\centering
\includegraphics[width=0.45\linewidth]{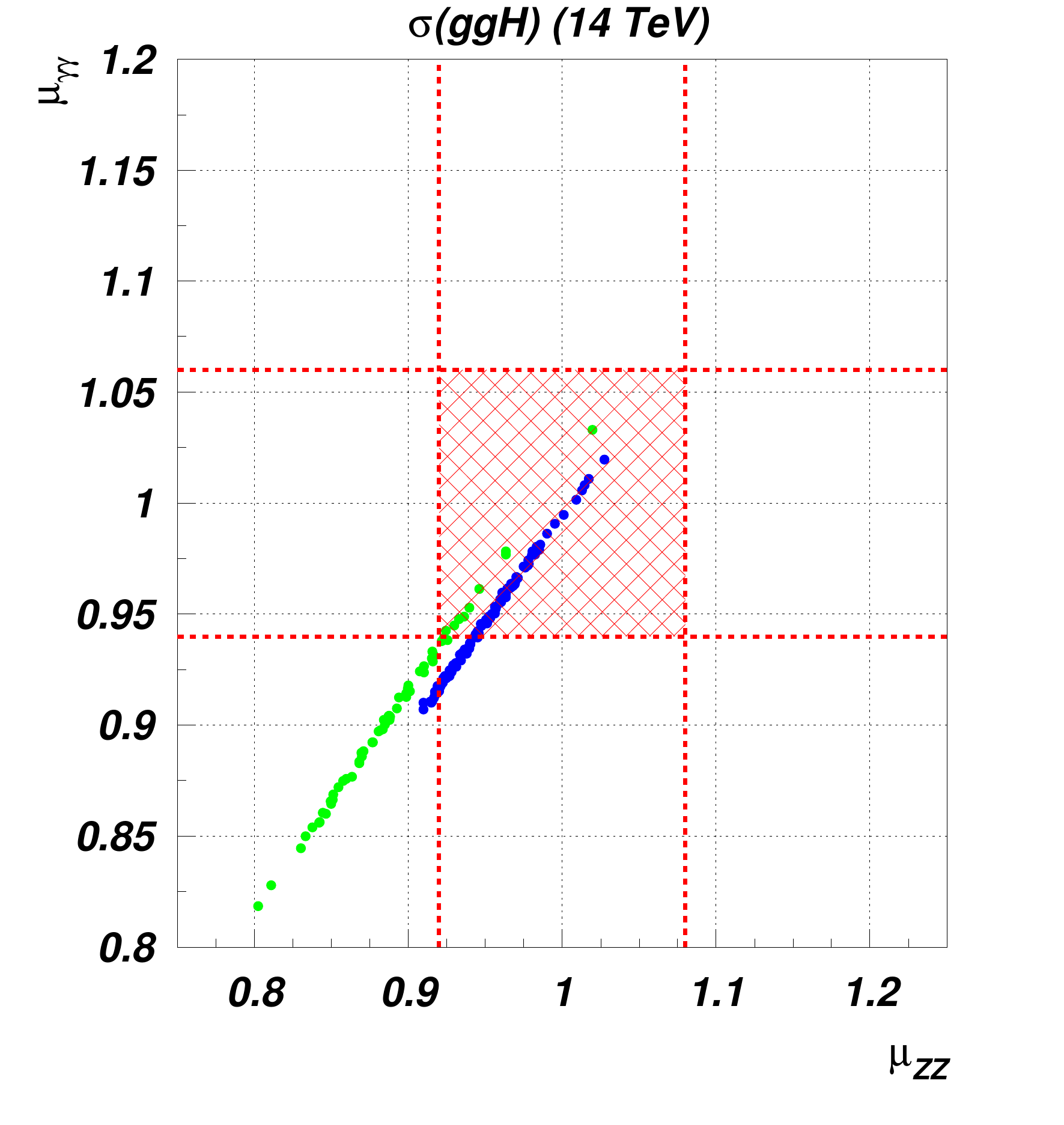}{}
\includegraphics[width=0.45\linewidth]{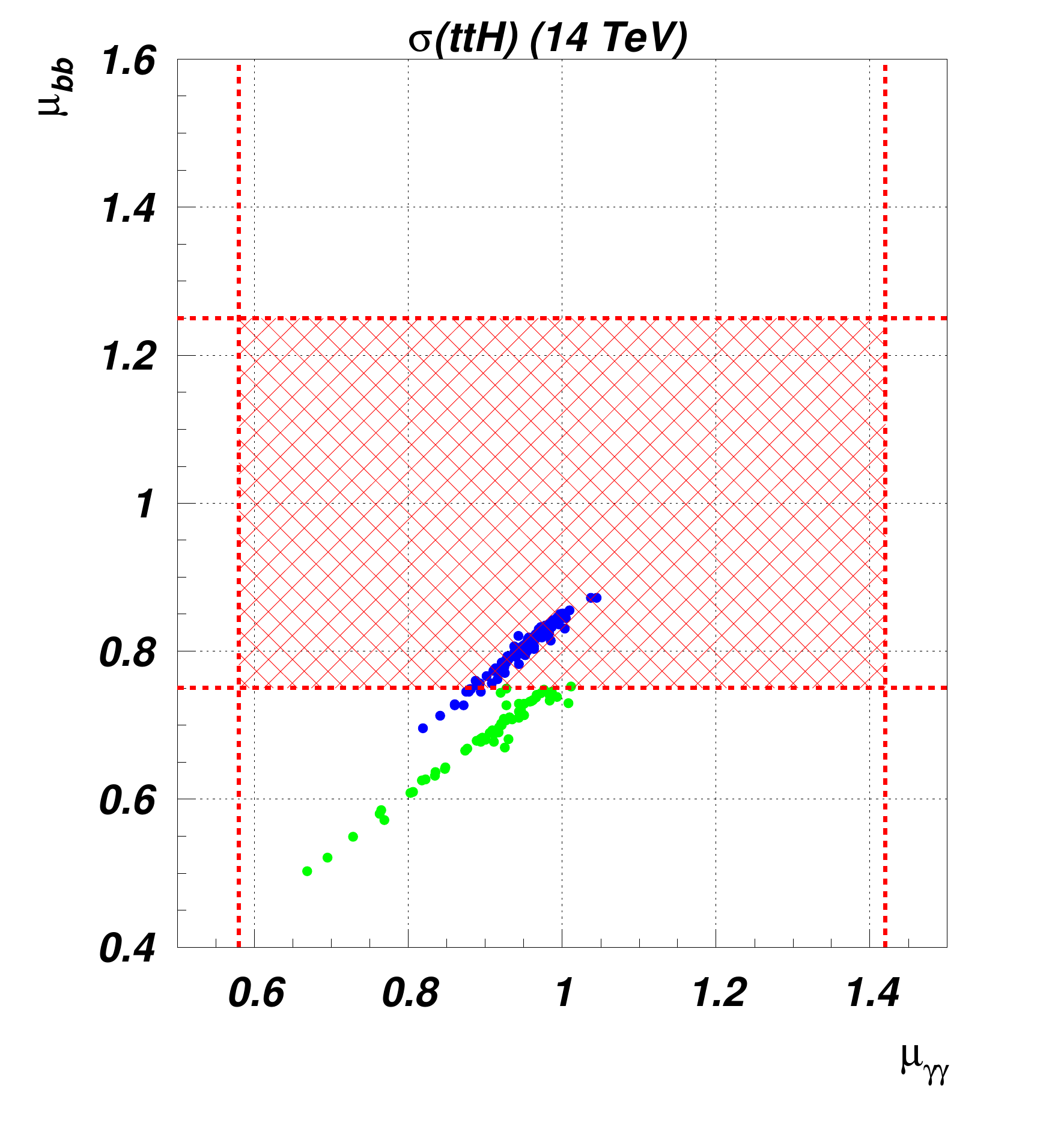}{}
\caption{Correlations among relevant $\mu_i$s evaluated at the LHC with 14 TeV and $300$ fb$^{-1}$ of luminosity, in 
$gg$ fusion (left frame) and associated production with $t\bar t$ (right frame).
Plots are for two 4DCHM benchmarks, with $f=800$ GeV and $g_\rho=2.5$ (green points) and $f=1000$ GeV and $g_\rho=2$ (blue points). The red shadowed areas represent the experimental precision limits around the SM expectations according to Tab.~\ref{tab:accLHC}.
}
\label{fig:4DCHMLHC}
\end{figure}

As a first step we consider the LHC capability to disentangle the 4DCHM from the SM by means of the measurement of the signal strengths $\mu_i$'s by comparing our predictions against the LHC experimental accuracies that can be found in Refs.~\cite{Peskin:2012we,Baer:2013cma}. In particular, 
we plot in Fig.~\ref{fig:4DCHMLHC}  two cases in which our parameter scans predict values of the signal strengths that lie outside the foreseen LHC experimental errors, that are reported for completeness in Tab.~\ref{tab:accLHC}~\footnote{The quoted sensitivities come from \cite{Peskin:2012we,Asner:2013psa,Baer:2013cma} where an analysis of ATLAS and CMS projections has been made allowing for an individual analysis of each Higgs production channel.}.
All the other predictions for the 4DCHM are inaccessible in terms of signal strength at the 14 TeV LHC with 300 fb$^{-1}$ of integrated luminosity and this reason, together with the ones already mentioned in the introduction, legitimate our analysis of the chosen model at a different machine, such as a future electron-positron collider.

\begin{table}[htb]
\captionsetup[subfloat]{labelformat=empty,position=top}
\centering
\begin{tabular}{||l|c||}
\hline
\hline
$ggH$		      & 14 TeV          \\
\hline
\hline
$ZZ$	    	     & 0.08		       	    \\
$\gamma\gamma$	     & 0.06		      	    \\
\hline
\hline
\end{tabular}
\quad
\quad
\begin{tabular}{||l|c||}
\hline
\hline
$ttH$		      & 14 TeV          \\
\hline
\hline
$b\bar b$	    	     & 0.25		       	    \\
$\gamma\gamma$	     & 0.42		      	    \\
\hline
\hline
\end{tabular}
\caption{Expected accuracies for cross section times BR  measurements for a 125 GeV Higgs boson produced via
$gg$ fusion (left frame) and in associated production with $t\bar t$ (right frame) as given in Refs.~\cite{Peskin:2012we,Baer:2013cma} at the
LHC with 14 TeV and 300 fb$^{-1}$ of luminosity.}
\label{tab:accLHC}
\end{table}

\newpage
\subsubsection{Higgs couplings analysis at $e^+e^-$ colliders in HS and VBF}
\label{Sec:VHWW}

We have shown in Sect.~\ref{Sec:decoup} that the exchange of sufficiently light new particles in the tree-level diagrams can affect Higgs production cross sections due to interference effects between SM and 4DCHM diagrams, so that the low energy approach cannot be sufficiently accurate in the processes that concern our analysis. Furthermore, as already intimated,
extra modifications to the various observables can also arise because of the presence of non-decoupled extra gauge bosons that give extra modifications (via mixing) to the couplings involved in the HS and VBF processes.
These coupling alterations can affect both the Higgs-vector-vector and vector-fermion-fermion couplings.
Lastly, loop-induced couplings, such as $H\gamma Z$, $H\gamma\gamma$ and/or $Hgg$, could also be affected by the presence of new fermions and, for the former, bosons running in the corresponding loop diagrams.
All these effects can therefore modify the signal strengths in a way that may be detectable with the 
experimental accuracies
expected at future electron-positron colliders.

\begin{figure}[htb]
\centering
\includegraphics[width=0.45\linewidth]{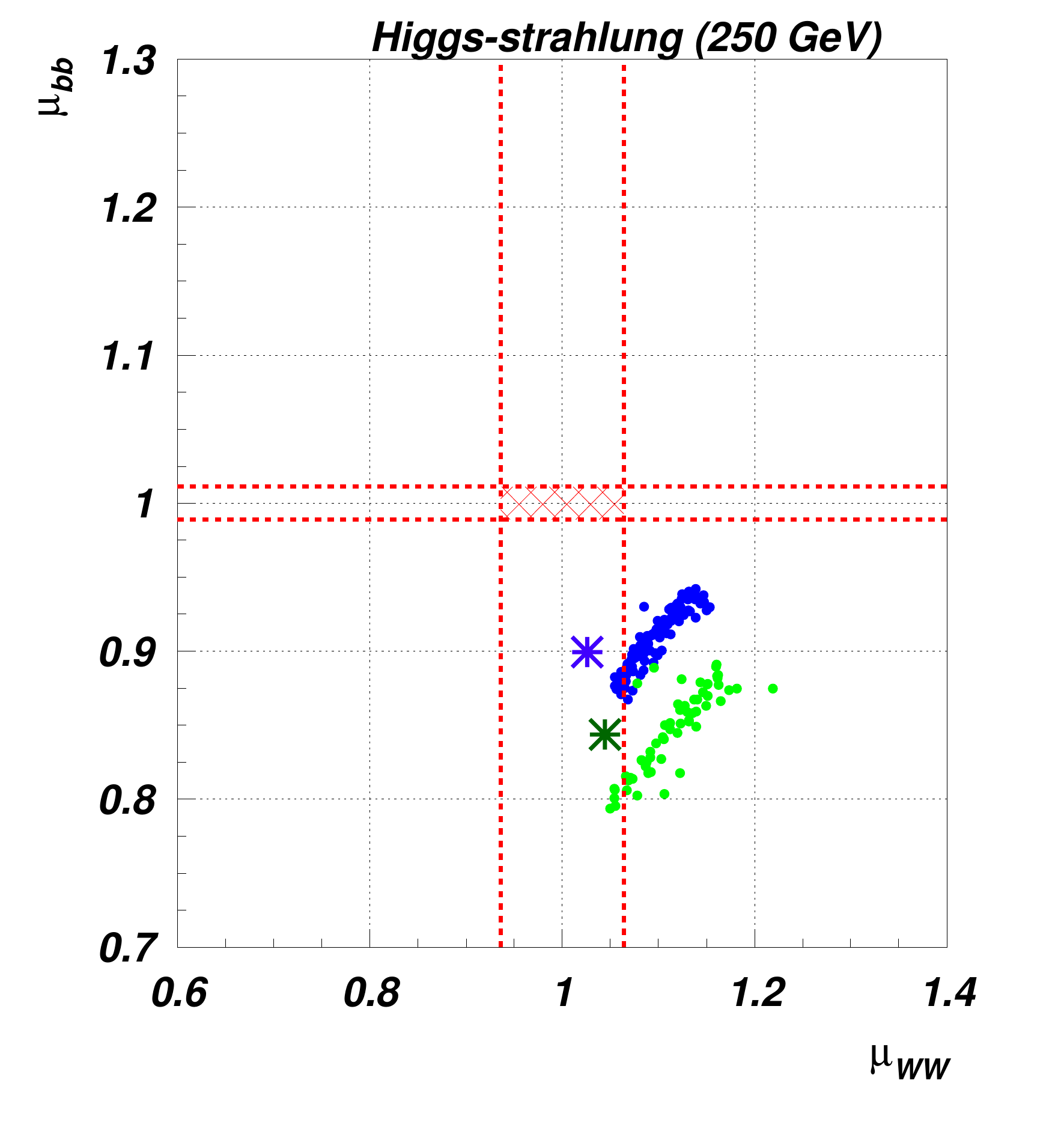}{}
\includegraphics[width=0.45\linewidth]{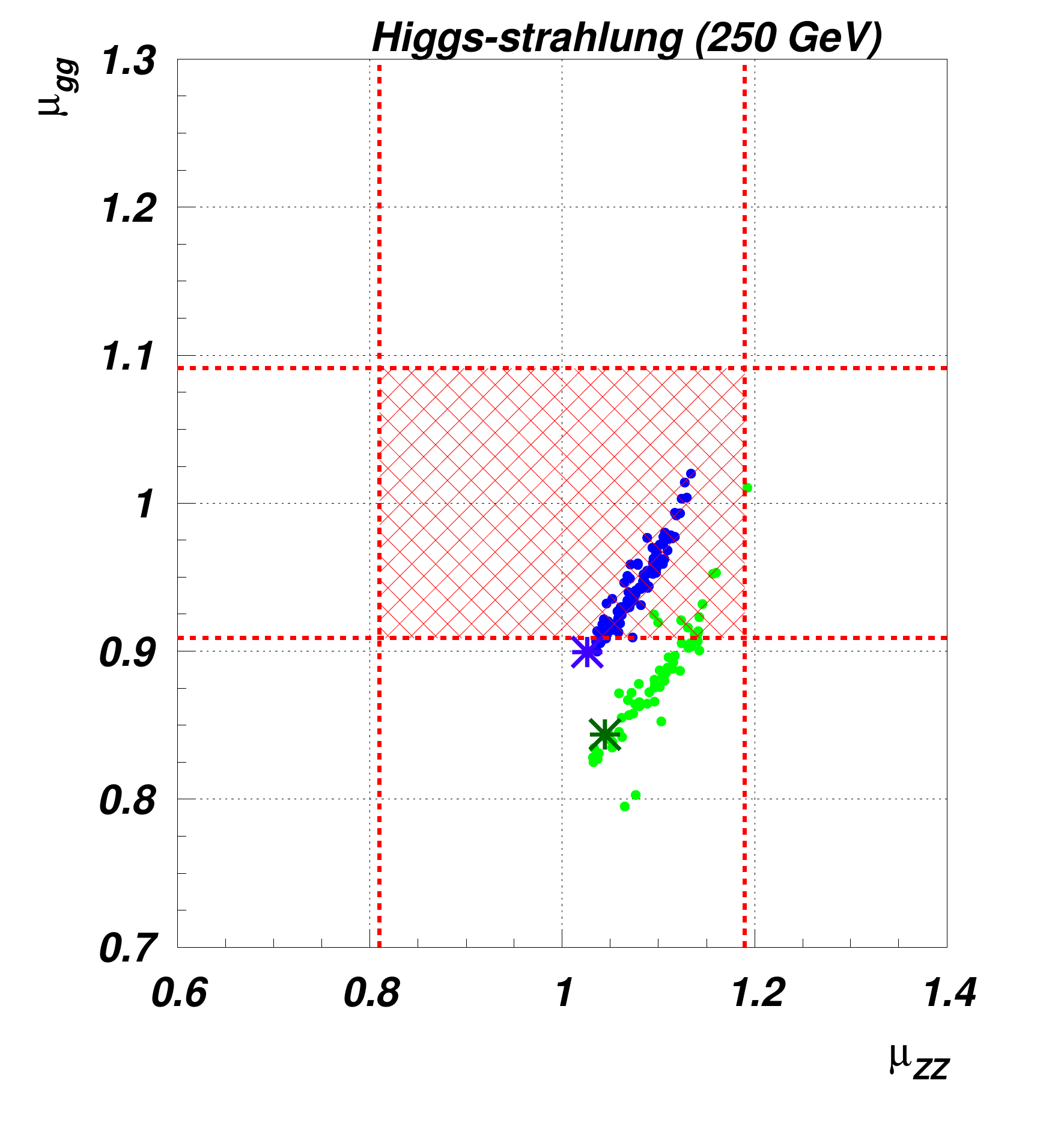}{}\\
\includegraphics[width=0.45\linewidth]{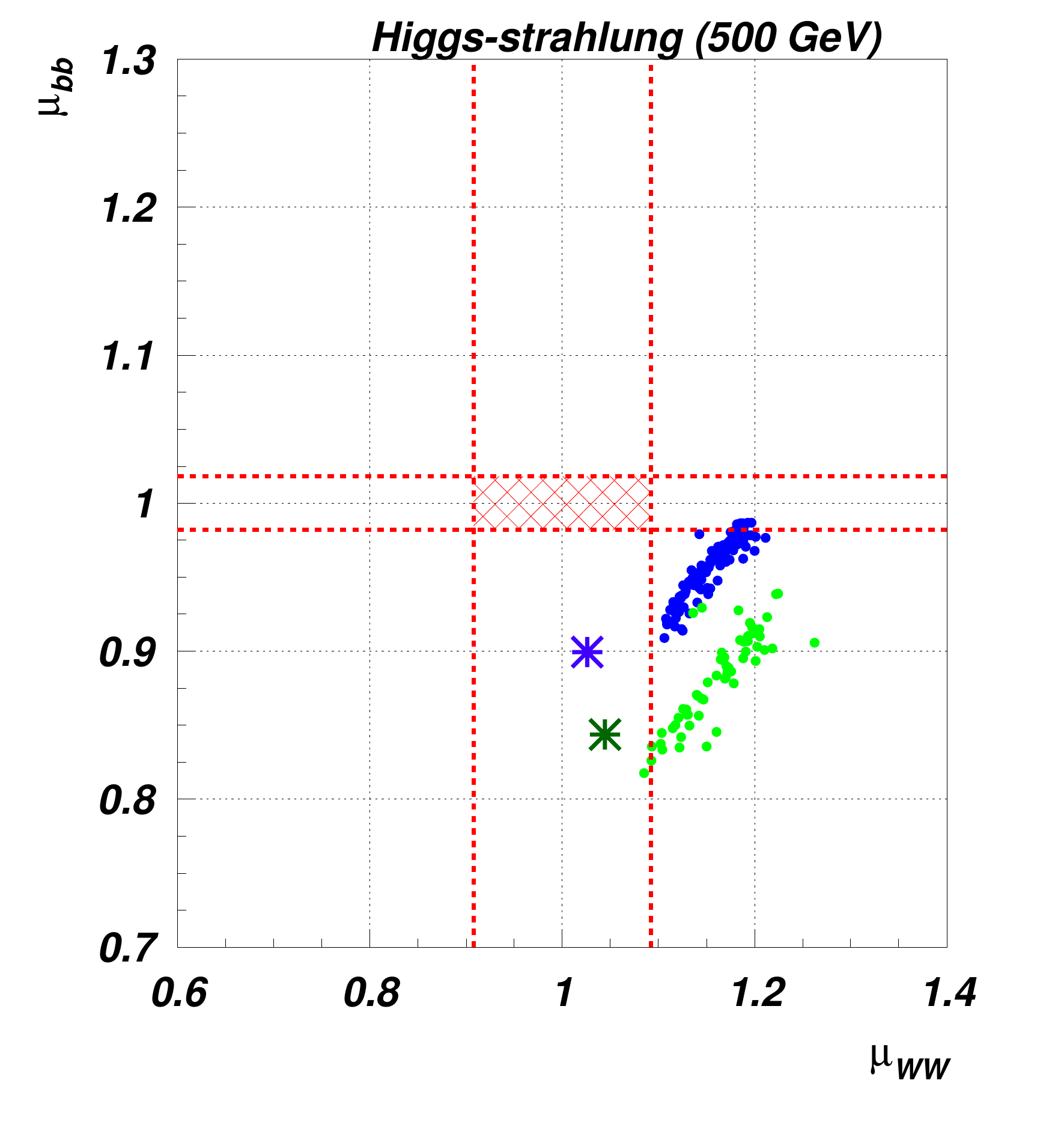}{}
\includegraphics[width=0.45\linewidth]{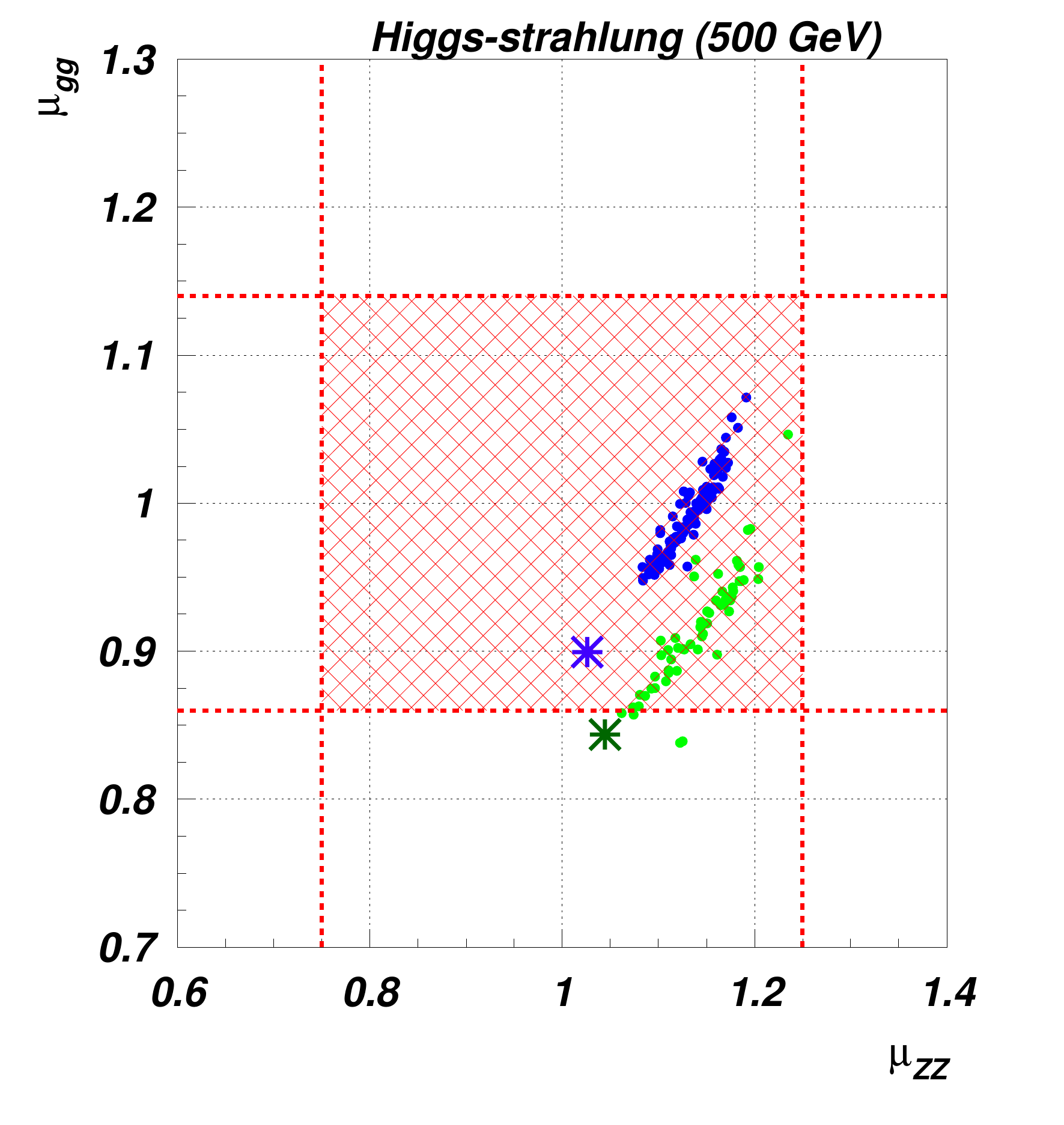}{}
\caption{Correlations among relevant $\mu_i$s evaluated at a future $e^+e^-$ collider
for two energy and luminosity stages, as detailed in the text, in the HS process. 
Plots are for two 4DCHM benchmarks, with  $f=800$ GeV and $g_\rho=2.5$ (green points) and $f=1000$ GeV $g_\rho=2$ (blue points). The red shadowed areas represent the precision limits around the SM expectations according to Tab.~\ref{tab:accLC}.
The asterisks represent the values obtained in the decoupling limit, whereby only the first effect of those mentioned in
Sect. IIIA is accounted for. 
}
\label{fig:4DCHMLC}
\end{figure}

\begin{figure}[htb]
\centering
\includegraphics[width=0.45\linewidth]{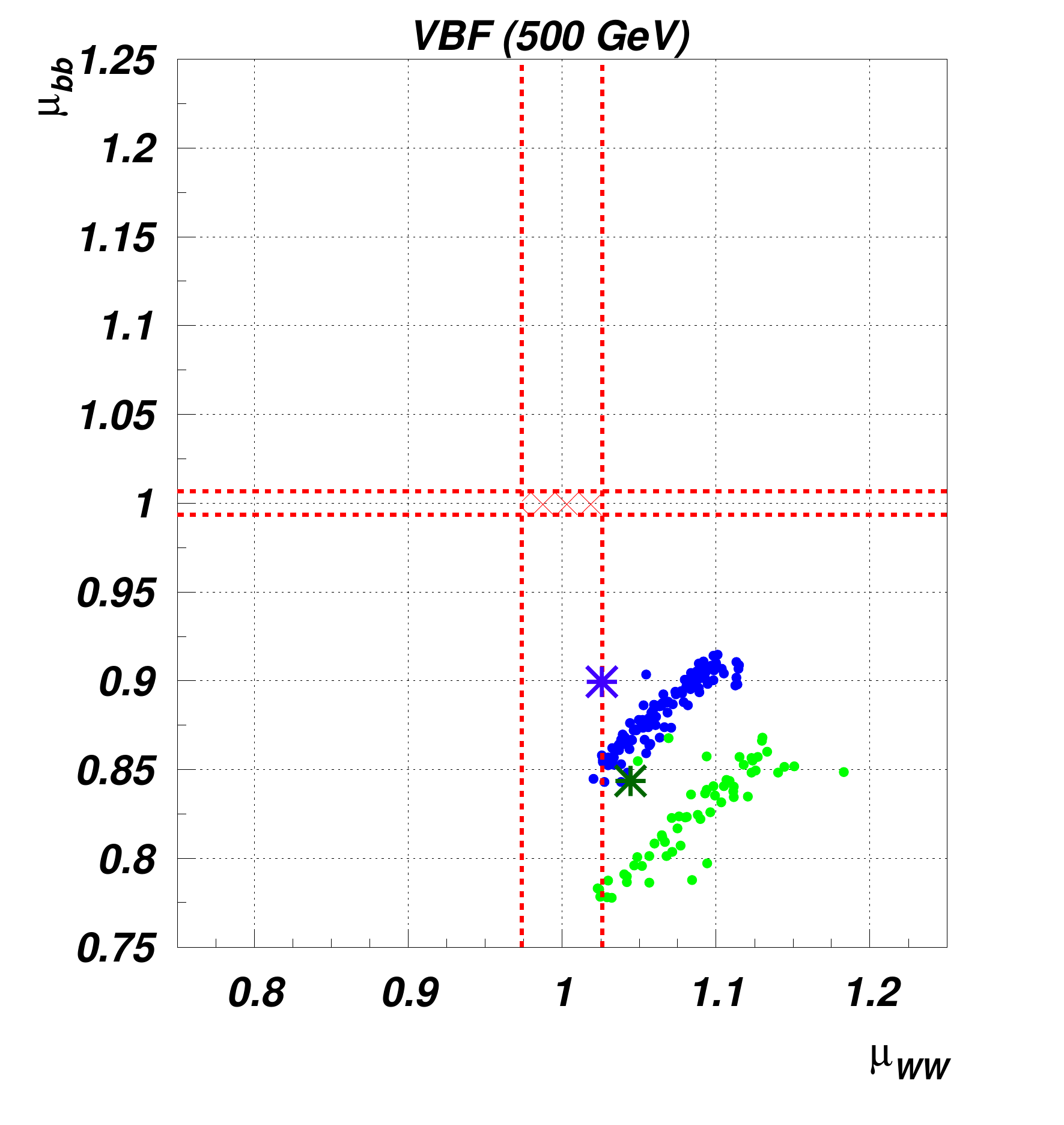}{}
\includegraphics[width=0.45\linewidth]{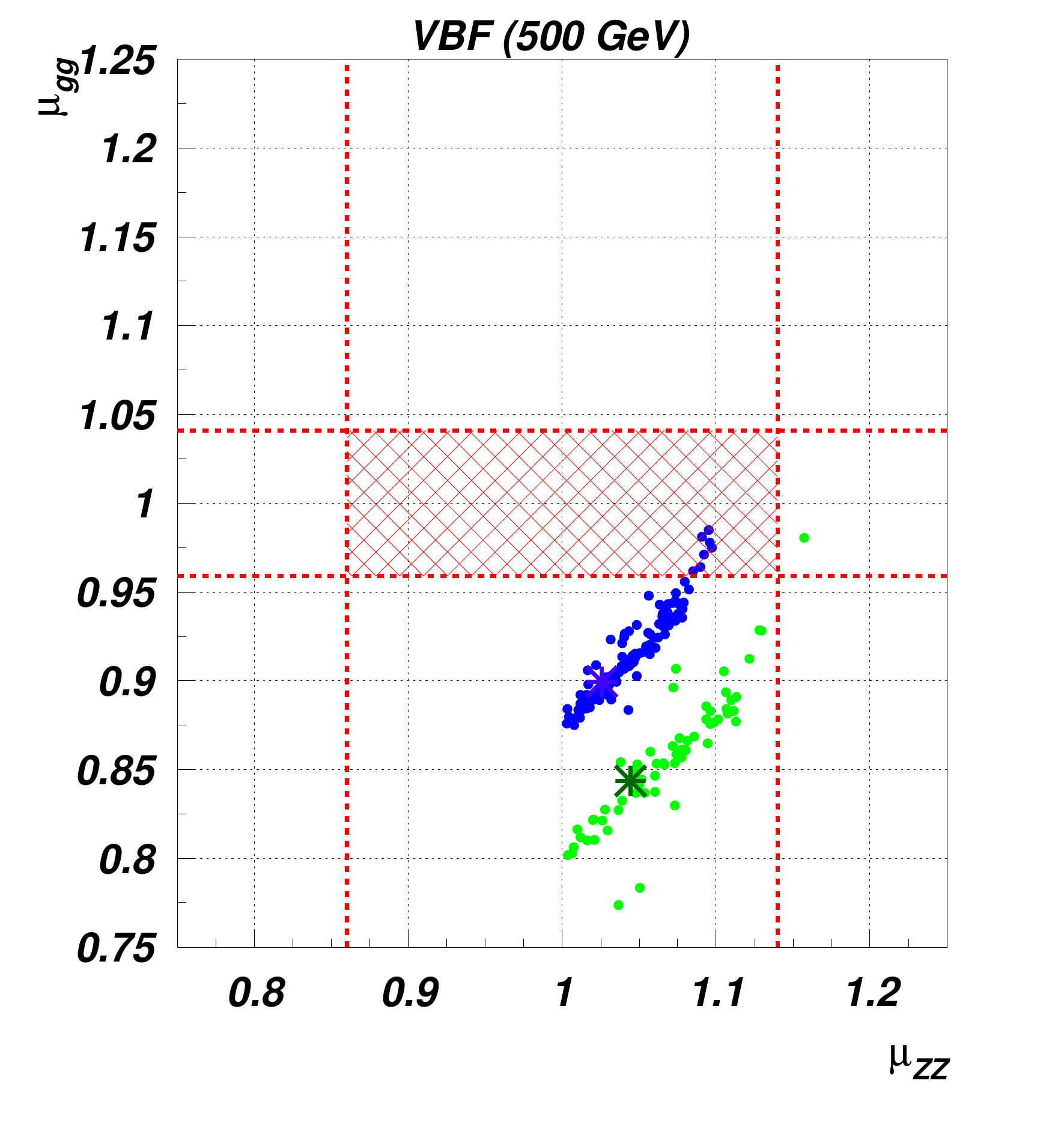}{}\\
\includegraphics[width=0.45\linewidth]{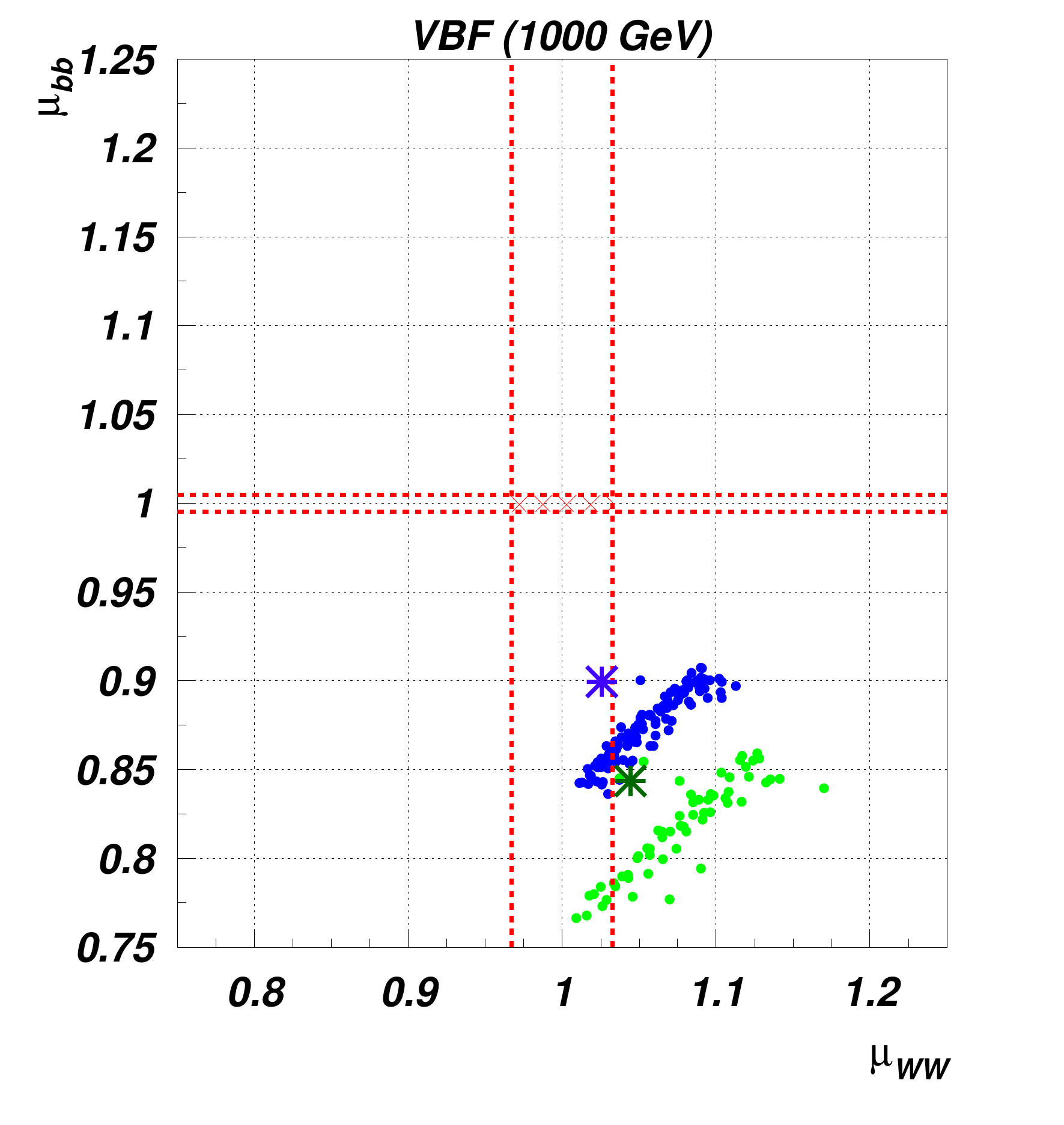}{}
\includegraphics[width=0.45\linewidth]{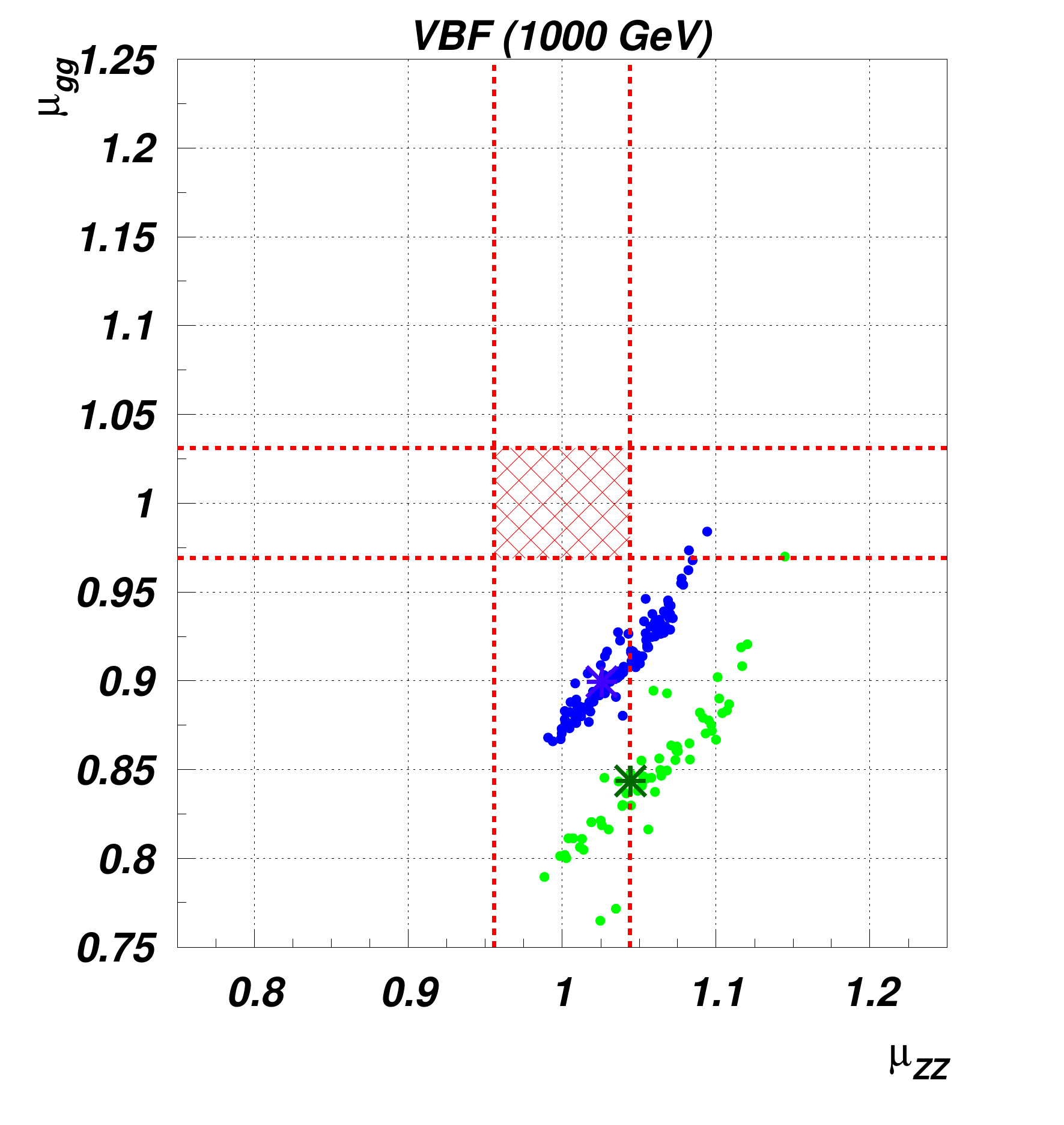}{}
\caption{Correlations among relevant $\mu_i$s evaluated at a future $e^+e^-$ collider
for two energy and luminosity stages, as detailed in the text, in the VBF process. 
Plots are for two 4DCHM benchmarks, with  $f=800$ GeV and $g_\rho=2.5$ (green points) and $f=1000$ GeV $g_\rho=2$ (blue points). The red shadowed areas represent the precision limits around the SM expectations according to Tab.~\ref{tab:accLC}.
The asterisks represent the values obtained in the decoupling limit, whereby only the first effect of those mentioned in
Sect. IIIA is accounted for.  
}
\label{fig:4DCHMLCVBF}
\end{figure}

Motivated by this additional reason, we present our results in terms of scatter plots for two of the three benchmarks already used in Sect.~\ref{Sec:decoup}: $f=800$ GeV, $g_\rho=2.5$ and $f=1000$ GeV, $g_\rho=2$.
We show the results of these scans in Figs.~\ref{fig:4DCHMLC} and  \ref{fig:4DCHMLCVBF} for HS and VBF production,
respectively.
From Fig.~\ref{fig:4DCHMLC} we notice that the deviations in the HS mode
from the case in which the full particle spectrum is not taken into account, represented by the stars, could modify the signal strengths for various channels. In some
cases these deviations are fully disentangleable while in others they are not, depending on where the scan points fall relative to the SM expectations and according to the corresponding experimental error bars for a particular signature.
In the case of VBF, see Fig. \ref{fig:4DCHMLCVBF}, the effects are still present although smaller due to the different topologies of the Feynman diagrams.
Altogether, though, it is clear the potential that future leptonic machines can offer in pinning down the possible composite nature of the 125 GeV scalar boson discovered at CERN by measuring its
`effective' couplings to essentially all SM matter and forces.

\begin{table}[htb]
\captionsetup[subfloat]{labelformat=empty,position=top}
\centering
\begin{tabular}{||l|c|c||}
\hline
\hline
HS		      & 250 GeV          & 500 GeV \\
\hline
\hline
$\sigma({ZH})$	     & 0.025		     &	            \\
$b\bar b$    	     & 0.011		     & 0.018	    \\
$WW$	  	     & 0.064		     & 0.092  	    \\
$ZZ$	    	     & 0.19		     & 0.25  	    \\
$\gamma\gamma$	     & 0.38		     & 0.38  	    \\
$gg$		     & 0.091		     & 0.14	    \\
\hline
\hline
\end{tabular}
\quad \quad
\begin{tabular}{||l|c|c|c||}
\hline
\hline
VBF		    &  250 GeV          & 500 GeV &   1000 GeV  \\
\hline
\hline
$b\bar b$    	     & 0.105		                   & 0.0066	    		&  0.0047				\\
$WW$	  	     & 			   		   & 0.026  	 		&  0.033				\\
$ZZ$	    	     & 			  		   & 0.082  			&  0.044				\\
$\gamma\gamma$	     & 			  		   & 0.26  			&  0.10				\\
$gg$		     & 			   	   & 0.041 			& 0.031	    \\
\hline
\hline
\end{tabular}
\caption{Expected accuracies for cross section times BR  measurements for a 125 GeV Higgs boson produced via
HS (left frame) and VBF (right frame) as given in Refs.~\cite{Peskin:2012we,Baer:2013cma} at a future $e^+e^-$ collider
for various energy and luminosity stages, as detailed in the text.}
\label{tab:accLC}
\end{table}


\newpage

\newpage

\subsubsection{The top Yukawa coupling from $e^+e^-\to t\bar tH$}
At the running energy stages of 500 and 1000 GeV one of the most important process to be analysed at a 
future $e^+e^-$ collider is
$e^+ e^-\rightarrow t\bar t H$\footnote{Note that the contribution of the Higgs-strahlung diagrams is negligible at the energies considered with respect to those depicted in Fig.~\ref{fig:feynHtt}, see \cite{Baer:2013cma}.}.
In this channel, beside the effects already seen in the HS channel due to the exchange of $s$-channel extra gauge bosons, mainly $Z_2$ and $Z_3$, there could also occur the exchange of $t'$s that, if light enough and/or produced resonantly, could affect this production process significantly. In fact,
in general composite Higgs models these particles are predicted to be relatively light, around the TeV scale (so as to avoid a large fine tuning: see, e.g., \cite{Redi:2012ha} for a review on this) which is precisely the region that is starting to be tested by the LHC experiments.
As already mentioned, in order to naively take into account the direct search limits of extra coloured quarks coming from the LHC, we have restricted our parameter scans by only allowing for masses of these states greater than 600 GeV. This does not intend to be a precise and accurate exclusion, which is beyond the scope 
of this work, rather an indication of what a realistic mass bound on these states (which depends non-trivially on different BRs, interference effects and other subtleties) could be, an approach that is indeed sufficient for the purposes our analysis.

\begin{figure}[htb]
\centering
\includegraphics[width=0.45\linewidth]{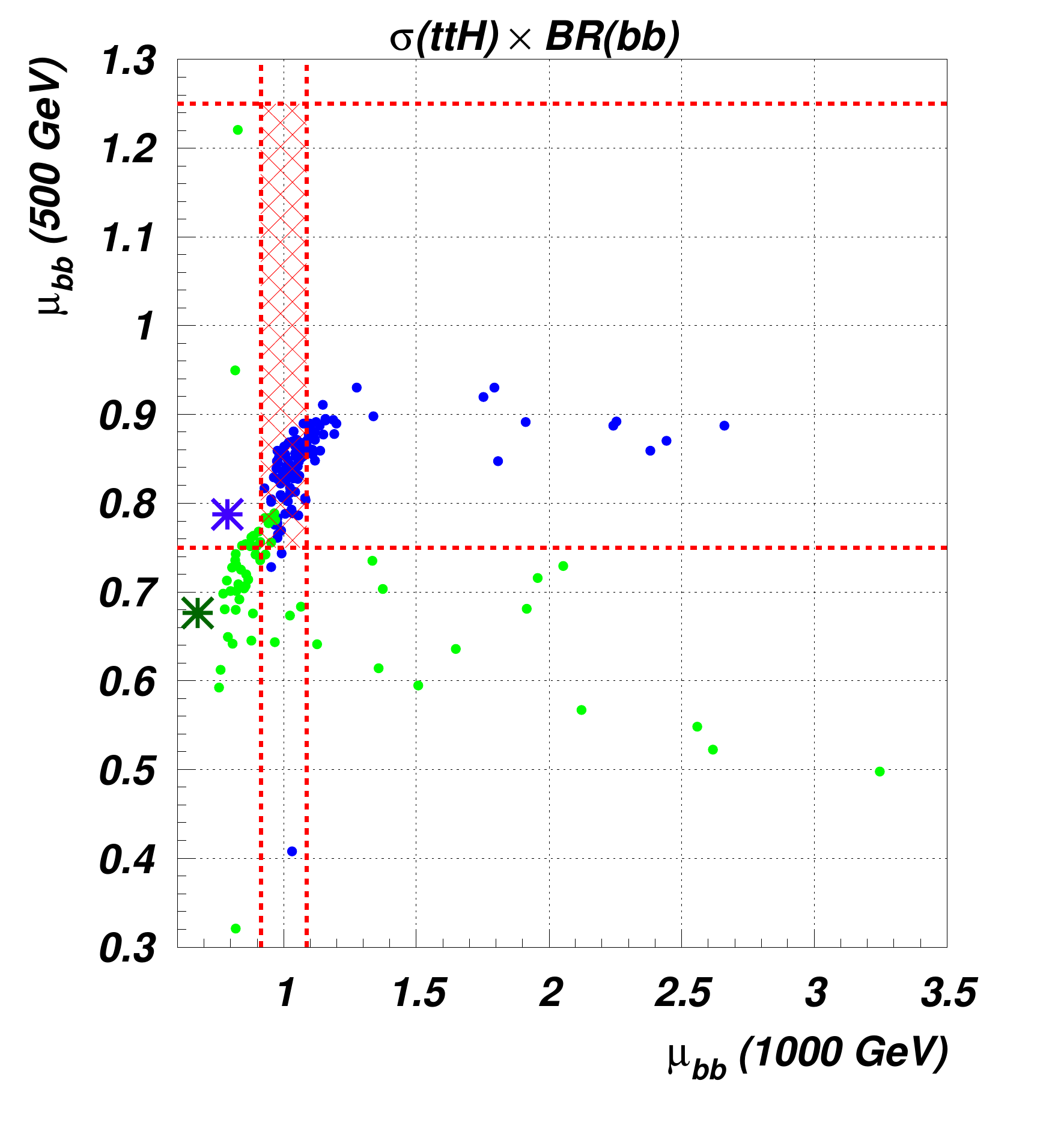}{}
\includegraphics[width=0.45\linewidth]{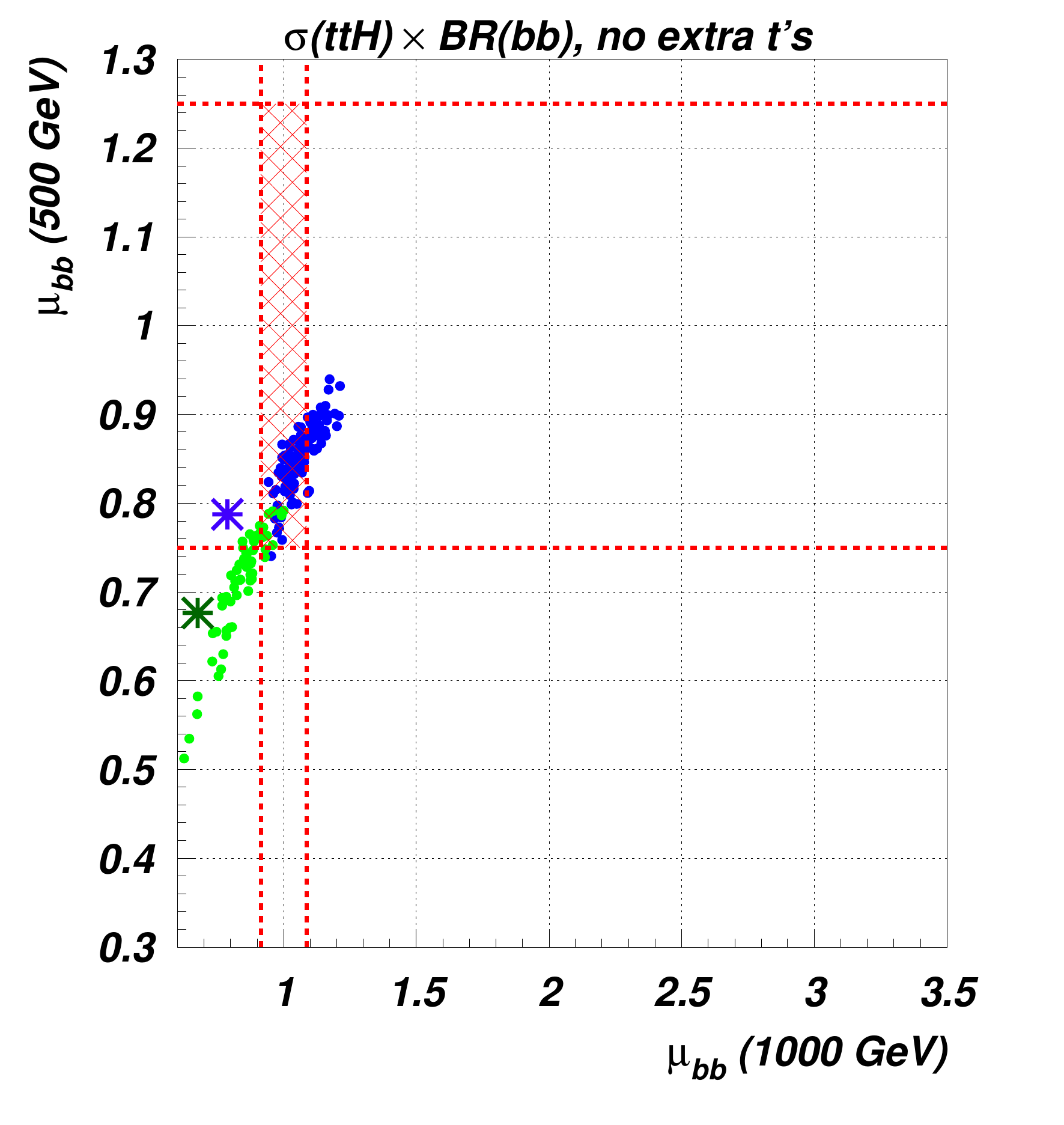}{}
\caption{Correlations among $\mu_{bb}$s evaluated at a future $e^+e^-$ collider
for two energy and luminosity stages, as detailed in the text, in the associated production with $t\bar t$,
with  the inclusion of $t'$ quarks (left frame) and without these (right frame). 
Plots are for two 4DCHM benchmarks, with $f=800$ GeV and $g_\rho=2.5$ (green points) and $f=1000$ GeV $g_\rho=2$ (blue points). The red shadowed areas represent the precision limits around the SM expectations according to Tab.~\ref{tab:acctopyuk}.
The asterisks represent the values obtained in the decoupling limit, whereby only the first effect of those mentioned in
Sect. IIIA is accounted for. 
}
\label{fig:TTH}
\end{figure}

Following again the guidance provided by Refs.~\cite{Peskin:2012we,Baer:2013cma}, we quote in Tab.~\ref{tab:acctopyuk} the expected accuracies, in correspondence to the chosen energies of a future $e^+e^-$ collider, of the $b \bar b$ signal strength (the most easily accessible one)
and we show the results of our scans in Fig.~\ref{fig:TTH}.
We present the results for the two benchmark points already used in the previous subsection in one case with the inclusion of $t'$ fermions while in the other we do so by excluding these extra states from the Higgs production process. From the comparison of the two panels of Fig.~\ref{fig:TTH} it is clear that the enhancement of the signal strength up to a value of 2.5 (or more) relative to the decoupling limit result
at $\sqrt s=1$ TeV is due to the exchange of $t'$ states with a mass smaller than $\sqrt{s}-m_{t}$ that can indeed be resonant in the subsequent production of a Higgs-top pair in the final state. In contrast, at 500 GeV, the
more moderate (yet still detectable) departure of the $\mu_{b\bar b}$ value from the the decoupling limit is due to 
non-resonant $Z'$ and $t'$ effects (from off-shell propagators and/or mixing), as at this energy (owing
to the aforementioned 600 GeV mass limit) no $t'$ mass can at the same time be larger than $m_t+m_H$ and smaller than  $\sqrt{s}-m_{t}$\footnote{Notice that, for $\sqrt s=500$ or 1000 GeV, $t'\bar t'$ production for $m_{t'}\ge600$ GeV is not possible, so that single $t'$ production followed by $t'\to tH$ could be
a discovery process of the heavy fermions or else an independent confirmation of a possible LHC signal of the latter.}.
Once again, the potential of a future $e^+e^-$ machine in accessing the 4DCHM and also assessing its finite mass effects is very significant also via this process.

\begin{table}[htb]
\captionsetup[subfloat]{labelformat=empty,position=top}
\centering
\begin{tabular}{||l|c|c||}
\hline
\hline
$ttH$		     & 500 GeV        	     & 1000 GeV 	\\
\hline
$b\bar b$    	     & 0.35		     & 0.087	   			\\
\hline
\hline
\end{tabular}
\caption{Expected accuracies for cross section times BR  measurements for a 125 GeV Higgs boson produced via
associated production with $t\bar t$  as given in Refs.~\cite{Peskin:2012we,Baer:2013cma} at a future $e^+e^-$ collider
for two energy and luminosity stages, as detailed in the text. Only the $b \bar b$ decay mode is considered.}
\label{tab:acctopyuk}
\end{table}

\subsubsection{The triple Higgs self-coupling}

As mentioned in Sect.~\ref{Sec:Model}, in composite Higgs models the Higgs potential is radiatively generated \emph{\`a la} Coleman-Weinberg \cite{Coleman:1973jx} and with the 4DCHM choice of the fermionic sector this turns out to be
UV finite.
From the effective potential and its derivatives at the minimum, one can extract the Higgs 
vacuum expectation value $v$,  its mass $m_H$ and  its triple self-coupling  $\lambda$.
At the leading order in the contribution of the gauge and fermionic loops we get
\begin{equation}
\lambda=\frac{3 m_H^2}{v}\frac{1-2 \frac{v^2}{f^2}}{\sqrt{1-\frac{v^2}{f^2}}} = \lambda_{\rm SM} \frac{1-2\xi}{\sqrt{1-\xi}},
\end{equation}
in agreement with \cite{Contino:2013gna}.

\begin{figure}[t!]
\centering
\includegraphics[width=0.50\linewidth]{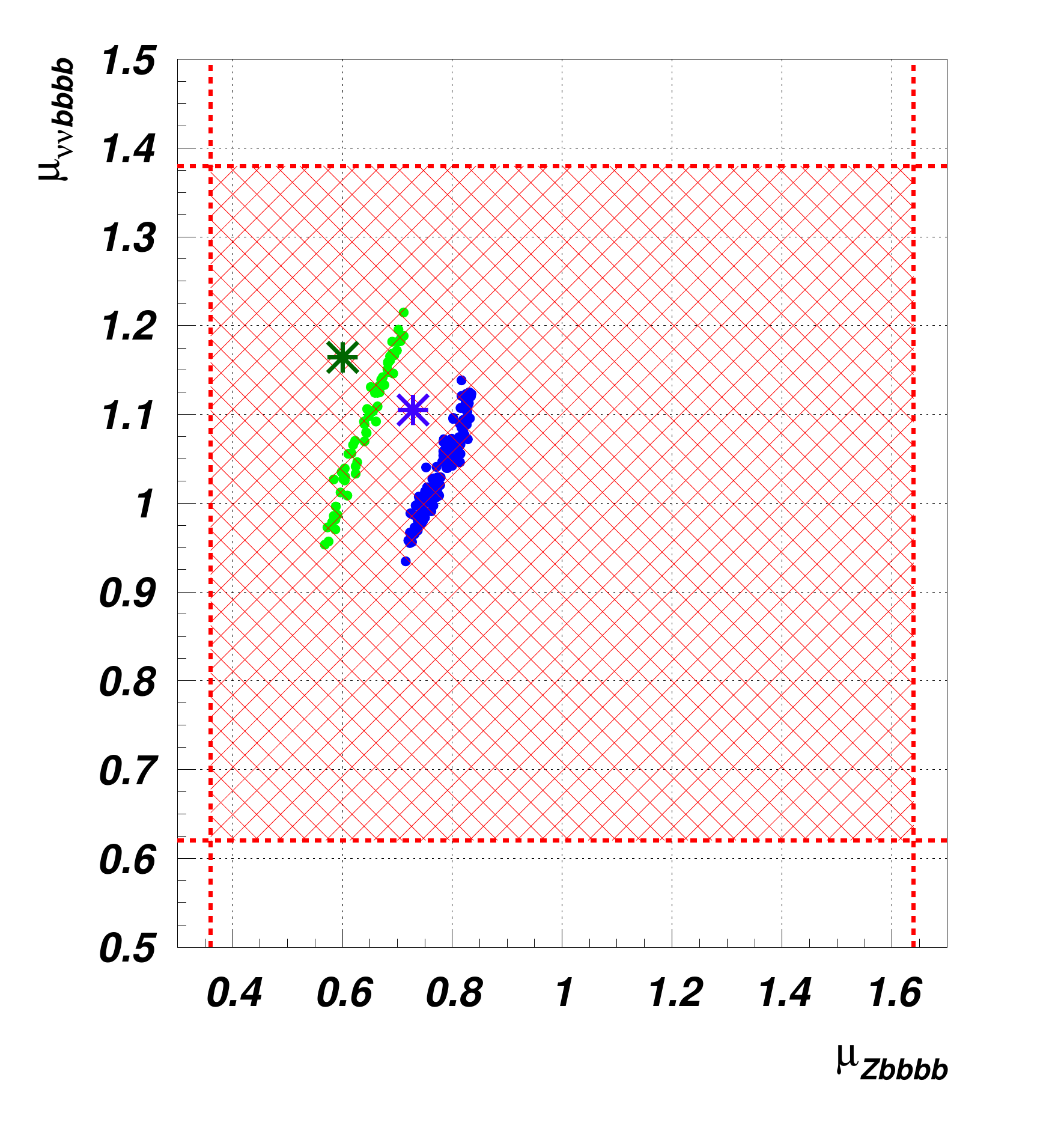}{}
\caption{Correlations among $\mu_{Zb\bar b b\bar b}$ (ratio of double Higgs production cross sections via HS and subsequent decay into $b \bar b$ final state)
and $\mu_{\nu_e\bar \nu_e b\bar b b\bar b}$ (ratio of double Higgs production cross sections via VBF and subsequent decay into $b \bar b$ final state)
evaluated at a future $e^+e^-$ collider
for two energy and luminosity stages separately for the two processes, as detailed in the text.
Plot is for two 4DCHM benchmarks, with $f=800$ GeV and $g_\rho=2.5$ (green points) and $f=1000$ GeV $g_\rho=2$ (blue points). The red shadowed areas represent the precision limits around the SM expectations according to Tab.~\ref{tab:accHH}.
The asterisks represent the values obtained in the decoupling limit, whereby only the first effect of those mentioned in
Sect. IIIA is accounted for. 
}
\label{fig:HH}
\end{figure}

This modified coupling intervenes in one of the two series of three Feynman diagrams of Fig.~\ref{fig:feynZHH}, which are, as mentioned, the ones concerning the HS and VBF double Higgs
production cross sections at a future $e^+e^-$ collider.
Our scatter plot in Fig.~\ref{fig:HH} illustrates that
sizable deviations in $\lambda$ are possible in the 4DCHM with respect to the SM, reflecting in a variation of production and/or decay rates at the level of several tens of percent.
However, accuracies in the corresponding measurements at a future electron-positron collider ought
to be reduced by a factor of two or so, in order to disentangle composite Higgs model
effects. 
This is in line with the  pursuit of the so-called ILC(LumUP) high luminosity scenario \cite{Asner:2013psa}, in the hope to decrease the expected error on such an observable down to a level comparable to the expected departures from the SM value.

\begin{table}[t!]
\captionsetup[subfloat]{labelformat=empty,position=top}
\centering
\begin{tabular}{||l|c|c||}
\hline
\hline
		     & $ZHH$ @ 500 GeV         & $\nu \bar \nu HH$ @ 1000 GeV 	\\
\hline
\hline
$b\bar b$    	     & 0.64		     & 0.38	   			\\
\hline
\hline
\end{tabular}
\caption{Expected accuracies for cross section times BR measurements for a 125 GeV Higgs boson produced in pair
via HS as given in Ref.~\cite{Baer:2013cma} at a future $e^+e^-$ collider
for two energy and luminosity stages, as detailed in the text. Only the $b \bar b$ decay mode of the Higgs boson is considered.}
\label{tab:accHH}
\end{table}

\section{Conclusions}
\label{Sec:Conclusions}
\noindent

In summary, we have used the 4DCHM to illustrate the potential of future $e^+e^-$ machines operating at energies between 250 and 1000 GeV, with appropriate luminosity options, in testing the salient features of composite Higgs models.
We have in particular shown that the approach of reducing the problem to a simple rescaling of SM production and decay rates and consequent extrapolations could be inaccurate in certain regions of the parameter space, as it fails to account for significant effects that can arise from mixing between SM and 4DCHM particles or interference effects between the two, whenever the mass spectra of extra gauge bosons or fermions present in these scenarios are allowed to exist around the
 compositeness scale, as indeed natural in these scenarios.  In fact, propagator effects of the additional heavy states present in composite Higgs models are often the most
visible signals. Such an approach instead naively integrates out all  heavy particles.
Clearly, the ensuing inaccuracies are the larger the smaller(higher) the new particle masses(collider energies) are.
In short, for mass spectra in either or both  the gauge and fermionic sectors of the 4DCHM  consistent with all current experimental and theoretical bounds, the effects studied here are tangible in most hallmark observables, such as Higgs production in single mode via HS, VBF, in association with  $t\bar t$ pairs as well as in 
(potentially) double Higgs production, both via HS and VBF.
Hence, our results indicate that this kind of studies must be based on a complete implementation of the relevant composite Higgs models, one example of which we make available for public use via the High Energy Physics Model Data-Base (HEPMDB) 
~\cite{Brooijmans:2012yi} at https://hepmdb.soton.ac.uk/ under the ``4DCHM'' name.


\section*{Acknowledgements} 

\noindent

DB and SM are financed in part through the NExT Institute. The work of GMP has been supported by the European Community's Seventh Framework Programme (FP7/2007-2013) under grant agreement n.~290605 (PSI-FELLOW/COFUND). DB and GMP would like to thank the Galileo Galilei Institute (GGI) in Florence for hospitality while part of this work was carried out.

\end{document}